\documentclass[
  amsmath,%
  amssymb,%
  pra,%
  aps%
  ]{revtex4-2}
\usepackage{lipsum}
\usepackage{graphicx}
\usepackage{bm}
\usepackage{hyperref}
\usepackage{xcolor}

\newcommand{\GM}[1]{{\color{black}{#1}}}

\newcommand{\bk}{\bar{\kappa}}

\newcommand{\hbk}{{\bar{\kappa}}}
\newcommand{\hk}{{\kappa}}

\hypersetup{
    colorlinks = true,
    linkcolor = blue,
    anchorcolor = blue,
    citecolor = red,
    filecolor = blue,
    urlcolor = blue
    }

\begin{document}

\preprint{APS/123-QED}

\title{Shortcuts to adiabaticity in harmonic traps: a quantum-classical analog}

\author{Vincent Hardel}
 \email{vincent.hardel@ipcms.unistra.fr}
\affiliation{%
    Universit\'e de Strasbourg, CNRS, Institut de Physique et Chimie des Mat\'eriaux de Strasbourg, UMR 7504, F-67000 Strasbourg, France
}%

\author{Giovanni Manfredi}
 \email{giovanni.manfredi@ipcms.unistra.fr}
 \affiliation{%
    Universit\'e de Strasbourg, CNRS, Institut de Physique et Chimie des Mat\'eriaux de Strasbourg, UMR 7504, F-67000 Strasbourg, France
}%
\author{Paul-Antoine Hervieux}
\email{paul-antoine.hervieux@ipcms.unistra.fr}
\affiliation{%
    Universit\'e de Strasbourg, CNRS, Institut de Physique et Chimie des Mat\'eriaux de Strasbourg, UMR 7504, F-67000 Strasbourg, France
}%
\author{R\'emi Goerlich}
\affiliation{
    Raymond \& Beverly Sackler School of Chemistry, Tel Aviv University, Tel Aviv 6997801, Israel
}

\date{\today}

\begin{abstract}
We present a new technique for efficiently transitioning a quantum system from an initial to a final stationary state in less time than is required by an adiabatic (quasi-static) process. Our approach makes use of Nelson's stochastic quantization, which represents the quantum system as a classical Brownian process. Thanks to this mathematical analogy, known protocols for classical overdamped  systems can be translated into quantum protocols. In particular, one can use classical methods to find optimal quantum protocols that minimize both the time duration and some other cost function to be freely specified.
We have applied this method to the time-dependent harmonic oscillator and tested it on two different cost functions: (i) the cumulative  energy of the system over time and (ii) the dynamical phase of the wavefunction.
In the latter case, it is possible to construct protocols that are ``adiabatically optimal", i.e., they minimize their distance from an adiabatic process for a given duration.
\end{abstract}

\maketitle

\section{Introduction}

Optimization problems play an important role in both classical and quantum physics, providing an elegant framework for describing natural phenomena. Concepts such as Fermat's principle and the stationary-action principle, which serve as foundational pillars of physics, are deeply rooted in optimization principles. Optimization techniques are also crucial for efficient resource utilization, improving theoretical models, and controlling and manipulating the state of a physical system.

The central question in optimization research revolves around bringing a system from an initial state to a desired final state while minimizing a certain quantity, known as the cost function. A very diverse array of systems can be explored through the lens of optimization, ranging from Kramers-like problems with double well potentials \cite{Martinez-Garaot2013}  to trapped ions \cite{Wineland1998}, cavity quantum electrodynamics \cite{Marr2003}, superconducting circuits \cite{Wendin2017}, spin-orbit coupling \cite{Khomitsky2012}, nitrogen-vacancy centers \cite{Song2016}, many-body and spin-chain models \cite{Bachmann2017}, and even metrology applications \cite{Pang2017}.
Various methods can be employed depending on the specific circumstances. Optimization techniques based on variational principles are often utilized, such as in the quantum brachistochrone problem \cite{Takahashi2013,Takahashi2015} or in the context of Bose-Einstein condensates trapped in harmonic potentials \cite{Li2016,Li2018,delCampo2011,delCampo2013}. Optimization methods based on optimal control theory have gained prominence in recent years \cite{Pinch1995,Kirk2012,Hegerfeldt2013,Sun2017,Zhou2017}, \GM{including more mathematical approaches such as Lyapunov and Krotov methods \cite{Layeghi2024,Morzhin2019}.
}

In the present work, we focus on the case of the time-dependent quantum harmonic oscillator, a system of paramount importance across multiple fields of physics. Specifically, we will examine protocols for manipulating the stiffness of the potential in order to efficiently transition the system from one steady-state to another in a time shorter than that required by an adiabatic (i.e., quasi-static) process. This concept is known as \emph{shortcut to adiabaticity} (STA) (for recent reviews, see \cite{Guery-Odelin2019a,Guery-Odelin2023a}).
While many of the basic ideas were put forward about two decades ago, with various approaches such as counterdiabatic driving \cite{Berry2009,Demirplak2005,Demirplak2008,An2016,delCampo2013,Keever2023}, inverse engineering \cite{Schmiedl2009b,Torrontegui2011}, scaling laws \cite{Muga2009}, and others \cite{Unanyan1997,Emmanouilidou2000,Couvert2008,Motzoi2009,Rezek2009,Salamon2009,Masuda2009,Demirplak2003} being introduced, the development of STA methods has gained much momentum in recent years \cite{Chen2010,Patra2017,Plata2021,Guery-Odelin2023b}.
\GM{Further, it should be mentioned that there exist lower bounds for the required time to transition from one quantum  state to another, known as \emph{quantum speed limits} (QSL) \cite{Okuyama2018,Poggi2021,Shanahan2018}. Our optimal protocols allow the transition in a duration that is situated in between the QSL and the adiabatic time, while, in addition, minimizing some other quantity of interest.
}

Here, we will employ an approach recently proposed for the classical overdamped dynamics of a Brownian particle confined in a harmonic trap and in contact with a heat bath at given temperature \cite{Rosales-Cabara2020a}. This method, based on a variational principle, allowed us to optimize the transfer from one thermal equilibrium to another, by minimizing both the duration of the transfer and the expended work.
The trade-off between duration and work could be modulated at will by tuning a single Lagrange multiplier.

The main purpose of the present work is to develop a quantum-classical analogy that allows us to exploit the aforementioned method in order to control and optimize the dynamics of a \emph{quantum} harmonic oscillator.
Indeed, analogies may serve as powerful tools in physics. For instance, the experiments conducted by Couder in 2005 \cite{Couder2005a} established a hydrodynamic analogy to the behavior of quantum particles, as described by the pilot wave theory introduced by de Broglie \cite{Broglie1927} and later developed by Bohm \cite{Bohm1952}. With this approach, it was possible to demonstrate the existence of quantum-like diffraction using a fully classical experimental setup \cite{Pucci2018}.
\GM{Some methods to obtain shortcuts to adiabaticity for both classical and quantum systems were developed in the past, particularly in the framework of counterdiabatic approaches \cite{Jarzynski2013,Deffner2014}. Our approach is different, inasmuch as it employs a strict mathematical analogy between a classical dissipative system and a quantum unitary evolution to translate classical protocols into quantum ones.

}

The analogy to be presented here is based on an alternative formulation of quantum mechanics originally due to Nelson \cite{Nelson1966a,Bacciagaluppi1999,Beyer2021}.
In Nelson's representation, the quantum evolution is governed by a first-order stochastic equation, supplemented by Schr\"odinger's equation for the wave guiding the trajectory in a manner similar to the Bohm-de Broglie theory. Nelson's stochastic equation bears a close resemblance to the Langevin equation that governs the overdamped motion of a classical Brownian particle, thus suggesting the potential for a fruitful quantum-classical analogy. Such an analogy has already been considered in recent years for an open quantum system \cite{Goerlich2021a}.
In the present work, we will demonstrate that this approach is particularly well-suited for solving optimization problems for closed quantum systems.

The objective of this paper is to devise a protocol that transitions the system from a given stationary state to another in the shortest possible time, while minimizing a specified cost function throughout the temporal evolution. In classical stochastic thermodynamics, it is common to minimize the work done on the system, which also corresponds to minimizing the dissipated heat \cite{Schmiedl2007a,Schmiedl2009b,Gomez-Marin2008a,Rosales-Cabara2020a}. However, for closed Hamiltonian systems, such work is simply equal to the difference between the final and the initial energies, and therefore it is not a pertinent quantity to minimize \cite{Chen2010}.

In the ensuing sections, we will outline a method -- based on Nelson's dynamics and the quantum-classical analogy mentioned above -- which allows us to minimize a generic cost function, usually written as the sum of the duration of the protocol plus a functional $F$ of the control parameters. As relevant examples, we will choose for $F$ either the cumulative energy of the system over time or the dynamical phase of the wavefunction. The latter case allows us to construct protocols that are ``adiabatically optimal", i.e. protocols that, for a given duration, minimize their distance (in a precise mathematical sense) from an adiabatic process.

In Section \ref{sec:1}, we will detail the basic features of the  classical analog of the time-dependent quantum harmonic oscillator using Nelson's stochastic formulation, illustrating the analogy with a simple numerical example. In Section \ref{sec:2}, we will show how this analogy can be applied to the quantum harmonic oscillator. We will also describe a general method to obtain a protocol that is optimal in regards of both its duration and another cost functional to be specified at will. In Sec \ref{sec:numerical} we will present numerical results for optimal protocols obtained using two different cost functions as illustrative examples. Conclusions and perspectives for future work will be detailed in Sec. \ref{sec:conclusion}.

\section{Quantum-classical analogy}
\label{sec:1}

\subsection{Fundamentals of the analogy}
\label{sec:1A}

A one-dimensional particle of mass $m$ trapped in a time-dependent harmonic potential obeys the Schr\"odinger equation
\begin{equation}\label{schrodinger_equation}
      i\hbar\frac{\partial}{\partial t}\psi(x,t) = \hat H(t) \psi \equiv \left(-\frac{\hbar^2}{2m}\frac{\partial^2}{\partial x^2} + \frac{1}{2}\kappa(t)x^2\right)\psi(x,t),
    \end{equation}
where $\hbar$ is Planck's constant, $\kappa(t)$ is the time-dependent stiffness of the potential, and $\psi(x,t)$ is the wavefunction of the system at time $t$ and position $x$.
The optimization procedure developed in this work consists in designing a protocol $\kappa(t)$ which brings the system from an initial stationary state $\psi_\mathrm{i}(x)$ at time $t_\mathrm{i}$ to a final (also stationary) state $\psi_\mathrm{f}(x)$ at time  $t_\mathrm{f}$, in the shortest possible time $\Delta t = t_\mathrm{f} - t_\mathrm{i}$,  while minimizing a given cost function \cite{Bulatov1998,Feldmann2000,Chen2010}.

The derivation of these optimal protocols will be based on Nelson's  formulation of quantum mechanics.
In this approach, similarly to the Bohm-de Broglie formalism, quantum particles are supposed to have a well-defined position $x(t)$ evolving in time.
Unlike the deterministic trajectories followed by quantum objects in the Bohm-de Broglie theory, Nelson's theory postulates that each trajectory obeys the stochastic differential equation
\begin{equation}\label{nelson_equation}
      \mathrm{d} x(t) = b(x(t),t)\mathrm{d}t + \sqrt{2D}\,\mathrm{d}W(t) ,
\end{equation}
where $b(x,t)$ is the deterministic drift velocity, $D = \hbar / 2m$ is the diffusion coefficient, and $\mathrm{d}W(t)$ is the Wiener increment for a Markov process: $\langle \mathrm{d} W(t) \rangle = 0$,  $\langle \mathrm{d}W(t) \,\mathrm{d}W(t')\rangle =  t-t'$.
The origin of the stochastic nature of the dynamic of quantum particles is postulated but not explained by Nelson's theory.
    The key point in Nelson's approach is to define the drift coefficient as
 \begin{equation}\label{b_nelson}
      b(x,t) = \frac{\hbar}{m}\frac{\partial S(x,t)}{\partial x}+\frac{\hbar}{2m}  \frac{\partial \ln\rho(x,t)}{\partial x},
 \end{equation}
 where $S(x,t)$ and $\rho(x,t)$ are respectively the phase and squared modulus of the wavefunction $\psi(x,t)$, expressed in polar form as $\psi=\sqrt{\rho}\,\exp(iS)$.
 This definition of $b(x,t)$ ensures that the probability distribution of a large ensemble of trajectories $x(t)$ obeying Nelson's equation (\ref{nelson_equation}) converges to the square-modulus of the wavefunction, following Born's rule \cite{Hardel2023}.
If the initial probabilistic distribution of the trajectories follows Born's rule at $t=0$, it will do so for all successive times $t>0$. Hence, Nelson's theory reproduces the same results as the standard quantum mechanics based on the Schr\"odinger equation.

For the ground state of the harmonic oscillator, the Gaussian form of the wavefunction yields a simple form for the drift term of Eq. (\ref{b_nelson}), as we now show.
Due to the quadratic nature of the Hamiltonian in Eq. (\ref{schrodinger_equation}), if the initial wavefunction $\psi_\mathrm{i}(x,t=0)$ is Gaussian, then it remains Gaussian for all times $t>0$, and can be written as
 \begin{equation}\label{psi_gaussian}
  \psi(x,t) = \frac{1}{\sqrt[4]{2\pi s(t)}}\exp\left[-\frac{x^2}{4s(t)} + i\alpha(t)x^2+i\beta(t)\right] ,
  \end{equation}
 where $s(t) = \langle x(t)^2 \rangle$ is the time-dependent variance of the density, while $\alpha(t)$ and $\beta(t)$ are the dynamical and geometrical phases of the wavefunction, respectively. For the wavefunction to be a solution of  Eq. (\ref{schrodinger_equation}), the time-dependent coefficients $\alpha(t)$ and $\beta(t)$  must satisfy the following relations:
     \begin{equation}\label{alpha}
      \alpha(t)=\frac{m}{4 \hbar} \frac{\dot{s}(t)}{s(t)}, \quad
 \dot  \beta(t) = - \frac{\hbar}{4ms(t)} ,
    \end{equation}
where the dot denotes differentiation with respect to the time $t$.
The variance $s(t)$ is related to the standard deviation $\sigma = \sqrt{2s}$, which must obey the following Ermakov equation
\cite{Lewis1967}:
 \begin{equation}\label{ermakov}
      \ddot{\sigma}(t) + \frac{\kappa(t)}{m}\sigma(t) = \frac{4D^2}{\sigma^3(t)}.
 \end{equation}
The coupled equations \eqref{alpha}-\eqref{ermakov} provide a full, exact solution of the Schr\"odinger equation in the Gaussian form \eqref{psi_gaussian}.

From Eqs. \eqref{b_nelson} and \eqref{psi_gaussian}, one obtains immediately Nelson's drift velocity:
    \begin{equation}\label{b_harmonic}
      b(x,t) = \frac{\hbar}{m}\left( 2\alpha(t) - \frac{1}{2s(t)} \right)x,
    \end{equation}
 which, importantly, is linear in $x$. Therefore, we can rewrite Nelson's stochastic equation \eqref{nelson_equation} as
 \begin{equation}\label{nelson_equation2}
      \mathrm{d} x(t) = \frac{\hbar}{m}\left( 2\alpha(t) - \frac{1}{2s(t)} \right) x(t) \,
      \mathrm{d}t + \sqrt{2D}\,\mathrm{d}W(t) .
\end{equation}

The above Nelson equation \eqref{nelson_equation2} bears a striking resemblance with the Langevin equation for a classical overdamped Brownian particle in a harmonic potential of stiffness $\bar{\kappa}(t)$  and same diffusion coefficient $D$, which we write here as:
 \begin{equation}\label{langevin_equation}
      \mathrm{d}x(t) = -\frac{\bar{\kappa}(t)}{\gamma}x(t)\mathrm{d}t + \sqrt{2D}\,\mathrm{d}W(t) ,
    \end{equation}
where $\gamma$ is the usual Stokes drag coefficient, which we keep for dimensional reasons in the classical equation, but will disappear in the quantum results.
The equations \eqref{nelson_equation2} and \eqref{langevin_equation} are identical if we define the classical stiffness as:
\begin{equation} \label{kappabar}
    \bar{\kappa}(t) = \gamma \frac{\hbar}{m}\left( 2\alpha(t) - \frac{1}{2s(t)} \right) .
\end{equation}
Hence, our physical analogy is based on the mathematical equivalence between Nelson's equation \eqref{nelson_equation2} for a quantum particle in a harmonic oscillator and the Langevin equation \eqref{langevin_equation} for a classical Brownian particle.
In addition, for the classical Langevin equation, the variance $s(t)$ obeys the following closed evolution equation \cite{Manoliu1979,gardiner2009stochastic}
    \begin{equation}\label{sdot}
      \frac{ \mathrm{d} s(t)}{ \mathrm{d}t} = \frac{2}{\gamma}[ D\gamma - \bar{\kappa}(t)s(t)].
    \end{equation}

To complete the analogy, we need to specify the relationship between the stiffness $\kappa(t)$ of the quantum oscillator and the the stiffness $\bar{\kappa}(t)$ appearing in the classical stochastic process.
Taking the time derivative of Eq. \eqref{kappabar} and using Eqs. \eqref{alpha} and \eqref{sdot}, we arrive, after some algebra, at the following expression for the quantum stiffness
    \begin{equation}\label{quantum_protocol}
      \kappa(t)=\frac{\hbar^{2}}{2 m s^{2}(t)}+\frac{m}{\gamma} \dot{\bar{\kappa}}(t)-\frac{m}{\gamma^{2}} \bar{\kappa}^{2}(t) ,
    \end{equation}
written in terms of the classical stiffness $\bar{\kappa}(t)$ and its time derivative. Equation \eqref{quantum_protocol} serves as a crucial link in establishing the quantum-classical analogy, acting as a bridge between the quantum system and its classical analog.

 Finally, from Eq. \eqref{sdot} it results that, at equilibrium: $\bar{\kappa}_\mathrm{eq} = D\gamma/s_\mathrm{eq}$ for the classical case.
 For the quantum case, from Eq. \eqref{quantum_protocol} we obtain
     \begin{equation}\label{equilibrium}
      \kappa_\mathrm{eq} = \frac{m}{\gamma^2} \bar{\kappa}_\mathrm{eq}^2 =  \frac{D^2m}{s_\mathrm{eq}^2},
    \end{equation}
    which does not depend on the classical parameter $\gamma$, as expected.

Our strategy will be to suggest a classical protocol $\bar{\kappa}(t)$ and use Eq. \eqref{quantum_protocol} to obtain the corresponding quantum protocol. By construction, the evolution of the variance $s(t)$ will be identical for both cases and given by Eq. \eqref{sdot}. Therefore, if we can devise a classical protocol that brings the variance from an initial equilibrium state with $s(t_\mathrm{i})=s_\mathrm{i}$ to a final equilibrium state with $s(t_{\rm f})=s_{\rm f}$, then the corresponding quantum protocol will do the same.

\GM{We further note that this method, being based on the overdamped equation \eqref{sdot}, will bring the variance $s(t)$ smoothly to the desired target value $s_{\rm f}$, for which $\dot s_{\rm f}=0$, after which it will remain there for all subsequent times. This is possible precisely thanks to the Nelson approach which, being governed by an overdamped stochastic equation, relaxes the system to the final equilibrium and guarantees that it does not depart from it. In other words, we have transformed our original quantum unitary evolution (which would display spurious oscillations when perturbed) into an equivalent dissipative classical evolution, which in contrast decays naturally to the desired target state. This can be done through the nontrivial transformation \eqref{quantum_protocol} from the physical quantum stiffness $\kappa(t)$ to an effective classical stiffness $\bar\kappa(t)$.
}

\subsection{Example: STEP protocol}
\label{sec:1B}

 To illustrate the quantum-classical analogy, we examine a sudden protocol (frequently referred to as STEP) which consists in an abrupt change of the classical  stiffness from $\bar{\kappa}_\mathrm{i}$ to $\bar{\kappa}_\mathrm{f}$. However, because of the presence of a first derivative in Eq. \eqref{quantum_protocol}, it is necessary to smooth out such STEP protocol. The smoothed STEP is then defined as follows \cite{Martinez-Tibaduiza2021}
  \begin{equation}\label{step}
      \bar{\kappa}(t) = \frac{\bar{\kappa}_\mathrm{f}+\bar{\kappa_\mathrm{i}}}{2} + \frac{\bar{\kappa}_\mathrm{f}-\bar{\kappa_\mathrm{i}}}{2}\tanh\left(\frac{t-\tau}{\epsilon}\right).
    \end{equation}
The classical stiffness is centered at $t=\tau$ and becomes steeper and steeper as $\epsilon \to 0$.
For simplicity, we have used units for which: $\kappa_\mathrm{i}= 2$, $\hbar=\gamma=1$, $m = 1/2$, so that the quantum diffusion coefficient is $D = \hbar/2m = 1$. In these units, time is measured in units of $2/\omega_\mathrm{i}$ and the variance in units of $\hbar/m\omega_i$, where $\omega_\mathrm{i}=\sqrt{\kappa_\mathrm{i}/m}$ is the initial angular frequency of the harmonic potential.
In the examples below, we have used $\bar{\kappa}_\mathrm{i} = 2$ and $\bar{\kappa}_\mathrm{f}=4$, which in virtue of Eq. \eqref{equilibrium} yields: $\kappa_\mathrm{i} = 2$ and $\kappa_\mathrm{f}=8$, and variances $s_\mathrm{i} = 0.5$ and $s_\mathrm{f} = 0.25$.

\begin{figure}[htbp]
      \centering
      \includegraphics[width=0.4\paperwidth]{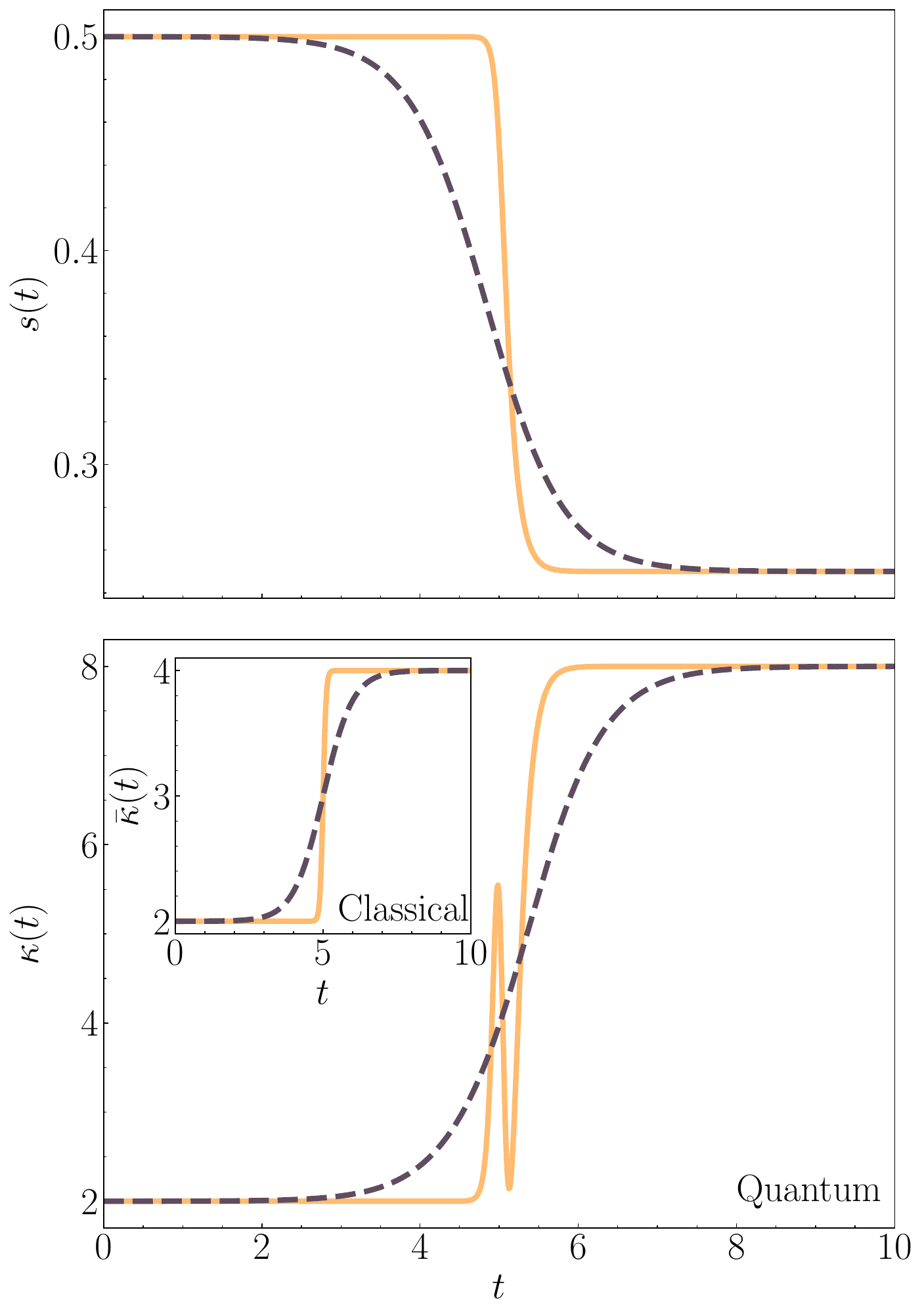}
      \caption{Top Panel: Evolution of the variance $s(t)$ as a function of time (measured in units of  $2\omega_\mathrm{i}^{-1}$), for $\epsilon =  1$ (black dashed lines) and $\epsilon = 0.1$ (orange solid lines). The variance decreases from $s_\mathrm{i} = 0.5$ to $s_\mathrm{f} = 0.25$, in units of $\hbar/m\omega_i$ as detailed in the main text. The larger value of $\epsilon$ correspond to the smoother protocol. Bottom panel: Time evolution of the quantum stiffness $\kappa(t)$ (main plot) associated to the classical STEP protocol $\bar{\kappa}(t)$ (inset), for the same values of $\epsilon$. When the classical STEP is smooth ($\epsilon =  1$, black dashed lines), the quantum protocol is also smooth and follows the same behavior as the classical one. In contrast, when the classical STEP protocol varies abruptly ($\epsilon =  0.1$, orange solid lines), an oscillation appears at mid-time in the quantum protocol. }\label{fig:smoothened_step}
\end{figure}

In Fig. \ref{fig:smoothened_step} (bottom panel), we show the classical and quantum protocols for two values of the width $\epsilon=0.1$ and $\epsilon=1$. For the smoother classical protocol ($\epsilon=1$, dashed black curve), the quantum protocol (shown in the inset) has a similar shape as its classical counterpart, although its initial and final values are different, in accordance with Eq. \eqref{equilibrium}. In contrast, the steeper classical protocol ($\epsilon=0.1$, solid orange curve) yields an oscillating quantum protocol (inset). These oscillations become stronger as $\epsilon \to 0$.
The time evolution of the variance $s(t)$ (top panel of Fig. \ref{fig:smoothened_step}) -- which, as stated above, is by construction the same for the classical and quantum cases -- shows that the variance of the system is smoothly brought from its initial value to its final value, even for the case ($\epsilon=0.1$) where the quantum protocol is strongly oscillating.

It is clear from Fig. \ref{fig:smoothened_step} that the steeper protocol achieves the transition more quickly, but let us try to quantify this speed-up more accurately. Classically, the relaxation time for a STEP protocol is given by the final stiffness $\bar \kappa_\mathrm{f}$, and reads as: $\tau_{\rm rel}=\gamma/\kappa_\mathrm{f}$. Rewriting this in terms of the quantum quantities, we get: $\tau_{\rm rel}= \sqrt{m/\kappa_\mathrm{f}} = \omega_\mathrm{f}^{-1}$, where $\gamma$, being a purely classical parameter, has naturally disappeared. Hence, the relaxation time is the inverse of the final oscillator frequency. In the present case, we have, in our units, $\tau_{\rm rel}= \omega_\mathrm{f}^{-1} = 0.25\,  (2\omega_\mathrm{i}^{-1})$.

Now, it is important to understand that, for the quantum oscillator (which is conservative), this is not really a relaxation time. If we apply a STEP protocol directly on the quantum stiffness $\kappa$, the quantum system will oscillate indefinitely, with no damping.  The standard way to implement the transition without oscillations would be to proceed adiabatically, which takes an infinite time. Hence, any quantum protocols, like those of Fig. \ref{fig:smoothened_step}, that take a finite time to complete, already do much better than the adiabatic one. In the figure, the slower protocol takes about $\approx 6\, (2\omega_\mathrm{i}^{-1})$ to achieve the transition, while  the faster protocol takes $\approx 0.75 \, (2\omega_\mathrm{i}^{-1})$. This speeding up is achieved through the special temporal profile of the quantum protocol $\kappa(t)$, which was obtained thanks to the quantum-classical analogy.

But the classical STEP protocol will never be able to go faster than the relaxation time $\tau_{\rm rel}= \omega_\mathrm{f}^{-1}$, which therefore constitutes a fundamental limit also for the quantum protocol, as the evolution of the variance is by construction the same for both. In the next section, we will develop a method to construct optimal protocols that break this limit, and allow relaxation on a timescale shorter than $\omega_\mathrm{f}^{-1}$. Therefore, these optimal protocols not only outperform the adiabatic process ($\tau_{\rm rel}= \infty$), but also do better than ``naive" protocols such as STEP, for which $\tau_{\rm rel} = \omega_\mathrm{f}^{-1}$.
In addition, they also minimize some other quantity of physical interest, such as the cumulative energy over time.

\section{Optimal quantum protocols}
\label{sec:2}
As mentioned in the preceding sections, our objective is to transition a quantum system from an initial state, $\psi_\mathrm{i}$, to a final state, $\psi_\mathrm{f}$, in the shortest possible time, $\Delta t$, while minimizing a certain cost function. For the time being, we keep this cost function as general as possible. The initial and final states are supposed to be the ground states in the respective harmonic potentials at  $t_\mathrm{i}$ and $t_\mathrm{f}$.
For a classical Brownian particle in a thermal bath, several methods already exist for finding the optimal protocol, especially when the cost function is the work done on the system~\cite{Rosales-Cabara2020a, Schmiedl2007b, Schmiedl2009b}. In the following, we will explore various cost functions and, for each of them, find the optimal protocol using the method based on the classical analogy developed in Sec. \ref{sec:1}.
Our approach is similar to the one used in our earlier work \cite{Rosales-Cabara2020a} and based on a variational principle.

In order to find the optimal protocol, we first need to define the functional to be minimized $J[\kappa,\dot \kappa]$, which is a function of the control parameter (stiffness of the oscillator) and its time derivative. This functional will be written as the sum of the total duration of the protocol, denoted $\Delta t[\kappa]$, plus another functional to be minimized, denoted $F[\kappa,\dot \kappa]$ (our cost function), the latter associated with a Lagrange multiplier $\lambda$. Then, the optimal protocol $\kappa_{\rm opt}$ is found as the solution of the Euler-Lagrange equation derived from the total functional $J$.
In practice, it will be necessary to add a third functional $G[\dot \kappa]$, and its Lagrange multiplier $\mu$, in order to ensure that the boundary conditions on $\kappa(t)$ are satisfied.

One has the choice to express the above functionals either in terms of the classical stiffness $\kappa$ or the quantum stiffness $\bar \kappa$, and then derive the corresponding Euler-Lagrange equations accordingly. In practice, as we shall see, it will be easier to express all functionals as a function of $\bar \kappa$ (and its time derivative) using Eq. \eqref{quantum_protocol}. For simplicity of notation, we will use the same symbols ($J$, $F$ and $G$) for the functionals, irrespective of their arguments.

Let us first express the time duration as a functional. We adopt the method developed in Ref. \cite{Rosales-Cabara2020a}, which consists in using the variance $s$ as an independent variable, instead of the time $t$ \footnote{This is possible only if $s(t)$ is a monotonic function of $t$.}. Each protocol can then be characterized by a trajectory in the stiffness-variance space $(\kappa,s)$. This allows us to express the protocols as a function of $s$ instead of $t$, so that we write: $\kappa(s) = {\kappa}(t(s))$ and $\bar{\kappa}(s) = {\bar{\kappa}}(t(s))$ (note that, for simplicity of notation, we use the same variable name for a function of $t$ and the corresponding function of $s$).
Using Eq. \eqref{sdot} one obtains for the time duration functional
  \begin{equation}\label{tf}
    \Delta t[{\bar \kappa}] = \frac{1}{2} \int_{s_\mathrm{i}}^{s_\mathrm{f}}\mathrm{d}s~ \frac{\gamma}{D\gamma - s{\bar{\kappa}}(s)}.
  \end{equation}
Note that the above functional depends on the classical protocol $\bar \kappa(s)$, and not the quantum one  $\kappa(s)$.

Since we are interested in the dynamics of the quantum system,
the yet-unspecified cost functional $F$ is usually expressed  in terms of $\kappa$,
the stiffness of the quantum harmonic oscillator.
However it is easy to express it in terms of $\bar \kappa$, by using Eq. (\ref{quantum_protocol}) rewritten
in the $s$-domain:
  \begin{equation}\label{quantum_protocol_s}
    {\kappa}(s) = \frac{\hbar^2}{2ms^2} + \frac{2m}{\gamma^2} [D\gamma - s{\bar{\kappa}}(s)]{\bar{\kappa}}'(s) - \frac{m}{\gamma^2}{\bar{\kappa}}^2(s),
  \end{equation}
where the prime denotes the derivative with respect to $s$. Then, we can write $F$ in integral form as:
$F[{\bar{\kappa}},{\bar{\kappa}}'] = \int_{s_\mathrm{i}}^{s_\mathrm{f}}\mathrm{d}s~f(s,{\bar{\kappa}}(s),{\bar{\kappa}}'(s))$, where $f$ is a function obtained by transforming the quantum stiffness to the classical stiffness using Eq. (\ref{quantum_protocol_s}).

The need for another functional $G[\kappa']$ results from the requirement to avoid strong gradients in $\kappa(s)$. As was noted in Ref.  \cite{Rosales-Cabara2020a}, without this term the resulting optimal protocol displays jumps, i.e. infinite gradients, at the initial and final times. Mathematically, this is because, in the absence of this term, the Euler-Lagrange equation is an algebraic one, so that one cannot fix the boundary conditions on the solution, and instead has to ``stitch" them artificially as jumps.
That was not an issue for the overdamped dynamics studied in  Ref. \cite{Rosales-Cabara2020a}, because a system with vanishing inertia remains at equilibrium when the stiffness is suddenly changed.
But here the situation is different, as the underlying problem is the standard Schr\"odinger equation, which does include inertia. Hence, if the boundary conditions are not satisfied at $t=t_\mathrm{f}$, then the system will continue to evolve and deviate from the target stationary state.
A way to ensure that boundary conditions are indeed satisfied is to render the Euler-Lagrange equation a second-order differential equation \cite{Boltyanski1960}. This can be achieved by adding the following functional
  \begin{equation}\label{G}
    G[{\bar{\kappa}}'] =  \int_{s_\mathrm{i}}^{s_\mathrm{f}}\mathrm{d}s~ \left | {\bar{\kappa}}'(s)\right |^2,
  \end{equation}
with the corresponding Lagrange multiplier $\mu$.
Note that we expressed $G[{\bar{\kappa}}']$ in terms of the classical stiffness, to be consistent with the other functionals $\Delta t$ and $F$. But limiting the gradient of $\bar{\kappa}(s)$ also limits the gradient of $\kappa(s)$, in virtue of Eq. \eqref{quantum_protocol_s}.
We also recall that the boundary conditions are those for which the variance is stationary at the boundaries, i.e. $\dot s_\mathrm{i,f} = 0$. From Eq.~(\ref{sdot}), this is equivalent to imposing that $\bar{\kappa}_\mathrm{i,f}\,s_\mathrm{i,f} = D\gamma$.

In summary, the total functional to be minimized is $J = \Delta t + \lambda F + \mu G$, or explicitly:
  \begin{equation}\label{J}
    J[{\bar{\kappa}},{\bar{\kappa}}'] = \frac{1}{2} \int_{s_\mathrm{i}}^{s_\mathrm{f}}\mathrm{d}s~ \frac{\gamma}{D\gamma - s{\bar{\kappa}}(s)} + \lambda\int_{s_\mathrm{i}}^{s_\mathrm{f}}\mathrm{d}s~f(s,{\bar{\kappa}}(s),{\bar{\kappa}}'(s)) +  \mu\int_{s_\mathrm{i}}^{s_\mathrm{f}}\mathrm{d}s~ \left | {\bar{\kappa}}'(s)\right |^2.
  \end{equation}
It is important to understand that $\lambda$ and $\mu$ play very different roles. While $\lambda$ corresponds to the weight given to the cost functional we want to  minimize on physical grounds, $\mu$ is present only to ensure that the equilibrium conditions are satisfied. Nevertheless, both Lagrange multipliers have an impact on the resulting optimal solution.
The Lagrangian associated to Eq. (\ref{J}) is
  \begin{equation}
    L[s,{\bar{\kappa}},{\bar{\kappa}}'] = \frac{\gamma}{D\gamma - s{\bar{\kappa}}} + \lambda f + \mu \left | {\bar{\kappa}}'\right |^2,
  \end{equation}
  where we removed the  factor $1/2$ from the first term, because it can be included in the Lagrange multipliers $\lambda$ and $\mu$. The optimal protocol  is obtained by solving the Euler-Lagrange equation associated with $L$, which reads as
  \begin{equation}
    2\mu{\bar{\kappa}}'' = \frac{\gamma s}{[D\gamma - s{\bar{\kappa}}(s)]^2} + \lambda \frac{\partial f}{\partial {\bar{\kappa}}} - \lambda \frac{\mathrm{d}}{\mathrm{d}s}\frac{\partial f}{\partial {\bar{\kappa}}'}.
  \end{equation}
  This equation is, as expected, a second-order differential equation for $\bar{\kappa}(s)$ and the two boundary conditions $\bar{\kappa}_\mathrm{i,f} = D\gamma / s_\mathrm{i,f}$ can thus be imposed.

\section{Optimization  results} \label{sec:numerical}

Some  protocols have already been studied in the past for the case where the cost function is the work done on the system~\cite{Schmiedl2007a}, i.e.: $F \equiv W =  \int_{t_\mathrm{i}}^{t_\mathrm{f}} \langle\psi|\partial_t \hat H(t)|\psi\rangle \mathrm{d} t$. It was shown that the optimal protocol is highly degenerate \cite{Schmiedl2009b}, with the minimal work simply corresponding to the difference between the final and initial energies of the system~\cite{Chen2010}: $W_\mathrm{opt} = E_\mathrm{f} - E_\mathrm{i}$.
In particular, for any protocol that satisfies the right boundary conditions -- so that the initial and final states are both stationary -- the work done on the system will be the optimal one, irrespective of the duration of the protocol \cite{Chen2010}.
This is not too surprising, as the quantum oscillator system is conservative, and therefore the energy expended to go from one stationary state to another should only depend on these states, and not on the path connecting them.
For these reasons, the work $W$ does not appear to be the pertinent cost functional to be minimized.
The advantage of the method described in the preceding section is that it allows us to select any cost functional, and find the corresponding protocol that minimizes it for a given duration $\Delta t$.
Here, we will consider two different functional forms of $F$ and find the optimal protocol for each of them.

In all the forthcoming simulations, we employ units in which $\hbar = \gamma= 1$, $m=0.5$, and then $D=1$.  The variance increases in time from $s_\mathrm{i} =1$ to $s_\mathrm{f} = 2$, so that the classical equilibria correspond to $\bar{\kappa}_\mathrm{i} = D\gamma/s_\mathrm{i} = 1$ and $\bar{\kappa}_\mathrm{f} = D\gamma/s_\mathrm{f} = 0.5$. The quantum equilibrium conditions are such that: $\kappa_\mathrm{i} = mD^2/s_\mathrm{i}^2 = 0.5$ and $\kappa_\mathrm{f} = mD^2/s_\mathrm{f}^2 = 0.125$. The classical relaxation time is then:
$\tau_{\rm rel}= \sqrt{m/\kappa_\mathrm{f}} =\omega_\mathrm{f}^{-1} = 2$.
All optimal protocols considered in the next two subsections are such that $\Delta t < \tau_{\rm rel}$ (except for one case where $\Delta t = 2.1$), confirming that they can outperform the STEP protocols, as discussed in Sec. \ref{sec:1B}.

Finally, in the Appendix \ref{annexe}, we will also treat the case of the classical optimal protocol developed in Ref. \cite{Rosales-Cabara2020a}, and discuss how it can be translated into an analog quantum protocol.

\subsection{Cumulative energy as cost function}\label{sec:energy-cost}

First, we take the  integral of the energy as the cost function, which, when divided by the total duration, corresponds to the time-averaged energy furnished to the system,  a quantity of clear physical interest, both theoretically and for experimental applications. As the wavefunction is supposed to be Gaussian at all times, see Eq. \eqref{psi_gaussian}, it is straightforward to write the energy as
\begin{equation}
      E(t) = \langle\psi|\hat H(t)|\psi\rangle = \frac{m}{4s(t)}\left(\frac{1}{2} \dot{s}^2(t) + \frac{2s^2(t)\kappa(t)}{m} + 2D^2 \right),
    \end{equation}
where $\hat H(t)$ is defined in Eq. \eqref{schrodinger_equation}.
We define the functional $F \equiv F_E$ as the integral of the energy. After changing variable from $t$ to $s$ and writing $\kappa(s)$ in terms of ${\bar{\kappa}}(s)$ using Eq. \eqref{quantum_protocol_s}, we obtain
    \begin{equation}
      F_E[{\bar{\kappa}},{\bar{\kappa}}'] = \frac{m}{4\gamma}\int_{s_\mathrm{i}}^{s_\mathrm{f}}\mathrm{d}s~ \left[  \frac{D\gamma - s{\bar{\kappa}}(s)}{s} + \frac{3D^2\gamma^2 - s^2{\bar{\kappa}}^2(s)}{s(D\gamma - s{\bar{\kappa}}(s))} + 2s{\bar{\kappa}}'(s)\right].
      \label{Fintegral}
    \end{equation}
The corresponding Lagrangian is then
    \begin{equation}\label{lagrangian_energy}
      L[s,{\bar{\kappa}},{\bar{\kappa}}'] = \frac{\gamma}{D\gamma-s{\bar{\kappa}}}+\frac{\lambda}{\gamma} \left[  \frac{D\gamma - s{\bar{\kappa}}}{s} + \frac{3D^2\gamma^2 - s^2{\bar{\kappa}}^2}{s(D\gamma - s{\bar{\kappa}})} + 2s{\bar{\kappa}}'\right] + \mu |{\bar{\kappa}}'|^2 ,
    \end{equation}
where the factor $m/4$ was absorbed in the Lagrange multiplier $\lambda$. The resulting Euler-Lagrange equation is a second-order nonlinear differential equation for $\bar \kappa(s)$, and the initial and final values of the classical protocol can be imposed according to Eq. \eqref{equilibrium}.
The Euler-Lagrange equation is
    \begin{equation}\label{euler_lagrange_energy}
      2\mu\gamma \, {\bar{\kappa}}'' = \frac{\gamma^2s + 3D^2\gamma^2 \lambda - s^2{\bar{\kappa}}^2 \lambda}{(D\gamma - s{\bar{\kappa}})^2} - \frac{2s\bar{\kappa} \lambda}{D\gamma - s{\bar{\kappa}}} - 3 \lambda,
    \end{equation}
and it can be  solved numerically using an iterative method such as Thomas's algorithm~\cite{Lipitakis2016}. Note that, had we not added the Lagrange multiplier  $\mu$,  Eq. \eqref{euler_lagrange_energy} would be an algebraic equation, and it would be possible to fix the correct boundary conditions.

\begin{figure}[htbp]
      \centering
      \includegraphics[width=0.4\paperwidth]{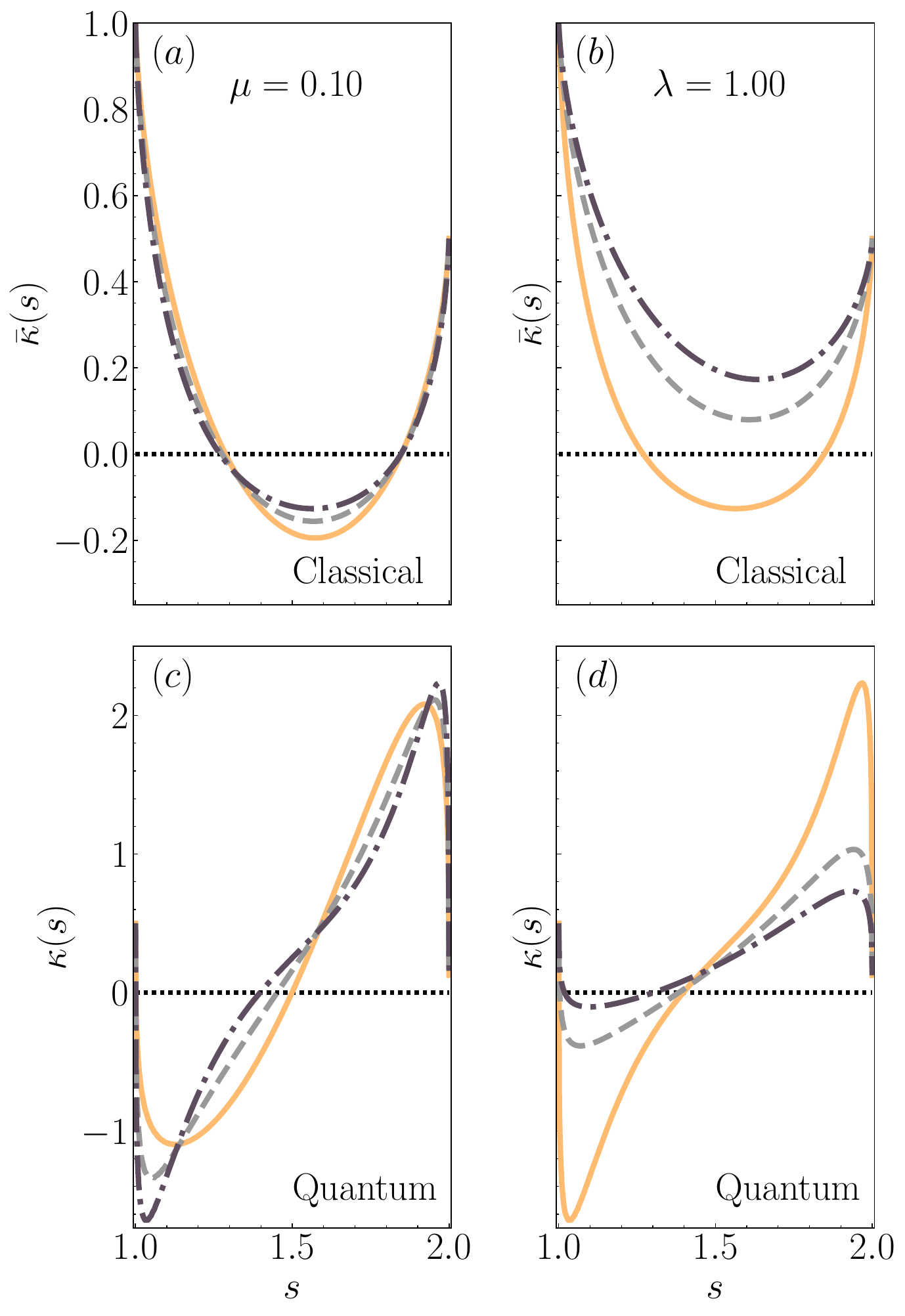}
      \caption{Energy-optimal protocols in the $s$-domain, where the variance increases from $s_\mathrm{i} =1$ to $s_\mathrm{f} = 2$.
      Left panels: Classical protocols $\hbk(s)$ (a) and quantum protocols $\kappa(s)$ (c),
      for a fixed value of $\mu = 0.10$, and  $\lambda = 0.01$ (solid orange lines), $0.10$ (dashed grey lines) and $1.00$ (dot-dashed black lines).
      Right panels: Classical protocols $\hbk(s)$ (b) and quantum protocols $\kappa(s)$ (d)
      for a fixed value of $\lambda =1$, and  $\mu = 0.1$ (solid orange lines), $0.50$ (dashed grey lines) and $1.00$ (dot-dashed black lines).
      Note that the quantum protocol can become negative for sufficiently large values of $\lambda$ or sufficiently small values of $\mu$.}
      \label{fig:energy_integral}
    \end{figure}

The solutions of Eq. (\ref{euler_lagrange_energy}) are represented in Fig.~\ref{fig:energy_integral}, for two distinct situations: a case where $\mu$ is kept fixed (left panels) and a case where $\lambda$ is kept fixed (right panels). For the first case ($\mu=0.1$ fixed), three solutions, corresponding to $\lambda = 0.01$, $0.10$, and $1.00$  are represented, thus showing the impact of varying the Lagrange multiplier associated with the cost function $F_E$.
The classical protocols are depicted in the upper panels of the figure, and the associated quantum protocols in the lower panels.
The equilibrium conditions at the initial and final times are fulfilled, as expected, and all protocols are continuous functions of the variance $s$.
Moreover, as $\mu$ is constant, the derivatives of each protocol at the boundaries of the $s$ domain are the same. Each classical protocol exhibits a minimal value, which decreases as $\lambda$ increases while its quantum associated protocol has both minimum and maximum values, whose amplitude increases as $\lambda$ increases. Hence,  both protocols can become negative, corresponding to a repulsive harmonic potential, when $\lambda$ is large enough. This feature, while notable, is not necessarily problematic and has been previously reported in the literature~\cite{Chen2010,Albay2020,Schmiedl2009b}.

The right panels of Fig.~\ref{fig:energy_integral} show  the solutions when $\lambda=1$ is kept constant while $\mu$ varies, taking the values $0.1$, $0.5$ and $1.00$. The shapes of the curves are similar to the preceding case, but it is notable that the larger $\mu$, the smaller the gradient of the protocol around $s_\mathrm{i}$ and $s_\mathrm{f}$, as expected because the functional $G$ limits the derivative of $\bar \kappa(s)$. Moreover, the range of values taken by the protocols increases as $\mu$ decreases, and they can become negative for small values of $\mu$. Caution should be taken for even smaller values of $\mu$, which lead to large values of the quantum protocol, both positive and negative. The same remark can also be made for very large values of $\lambda$.

\begin{figure}[htbp]
      \centering
      \includegraphics[width=0.4\paperwidth]{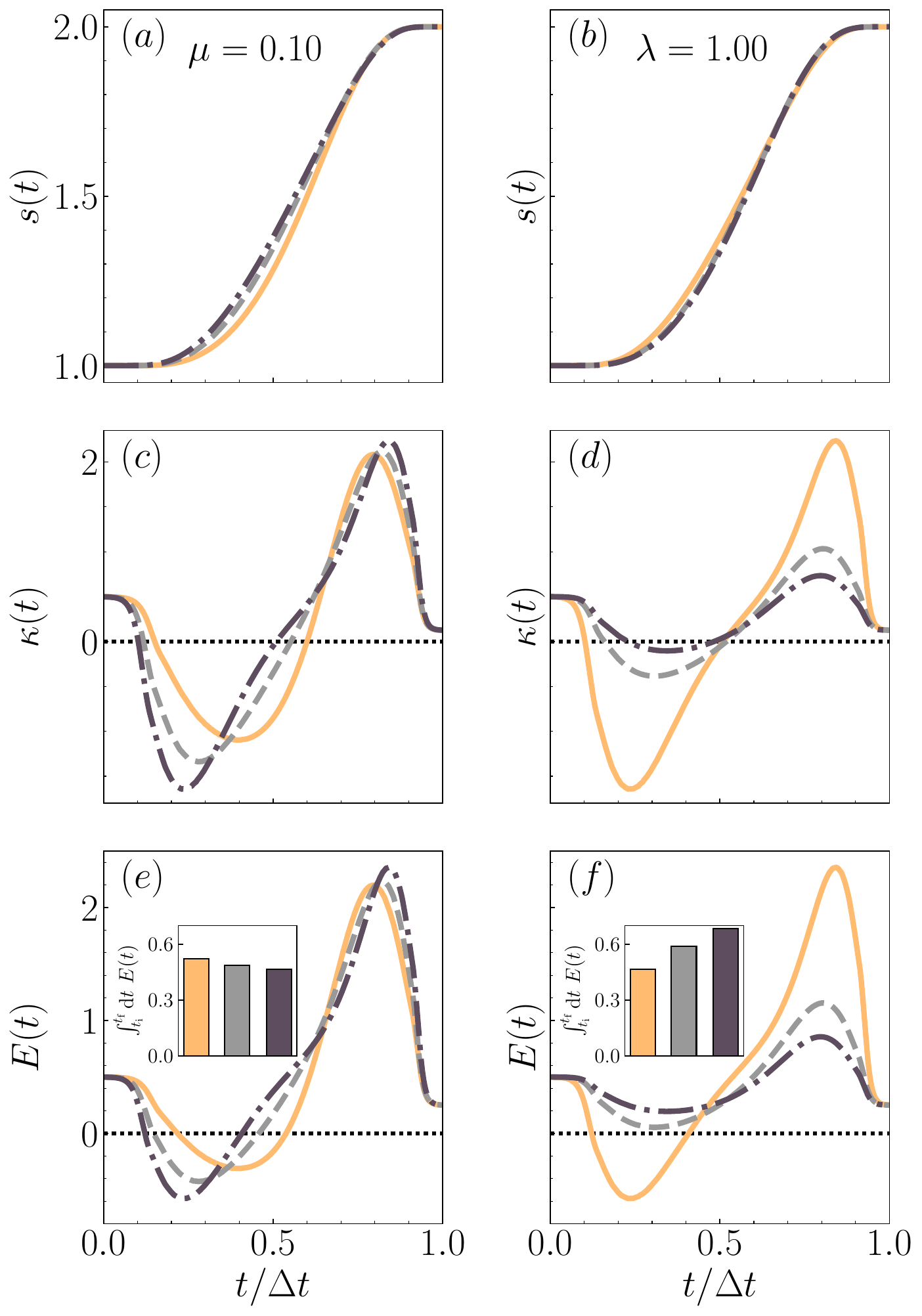}
      \caption{Energy-optimal protocols in the time domain, where the variance increases from $s_\mathrm{i} =1$ to $s_\mathrm{f} = 2$. The time is normalized to the total duration of the protocol $\Delta t$, which is different for each pair of Lagrange multipliers $(\lambda,\mu)$.
      Left panels: variance $s(t)$ (a), quantum protocols $\kappa(t)$ (c), and instantaneous energy $E(t)$ (e), for a fixed value of $\mu = 0.10$, and  $\lambda = 0.01$ (solid orange lines), $0.50$ (dashed grey lines) and $1.00$ (dot-dashed black lines). These values correspond to time durations: $\Delta t = 0.95$, $0.87$ and $0.82$.
      Right panels: variance $s(t)$ (b), quantum protocols $\kappa(t)$ (d), and instantaneous energy $E(t)$ (f), for a fixed value of $\lambda =1$, and  $\mu = 0.1$ (solid orange lines), $0.50$ (dashed grey lines) and $1.00$ (dot-dashed black lines).
      These values correspond to time durations: $\Delta t = 0.82$, $1.27$ and $1.58$.
      Note that, although the various protocols are quite different, the behavior of the variance is almost identical when time is scaled to the total duration $\Delta t$. The boundary values of the quantum protocols correspond to the conditions: $s_\mathrm{i,f}^2 \kappa_\mathrm{i,f} = mD^2$. The energy $E(t)$ is quite similar to the quantum stiffness $\kappa(t)$, indicating that the potential energy is the predominant contribution to the total energy of the system.
      The insets in (e) and (f) show the cumulative energy over time for the three protocols represented in the same panels.
}
      \label{fig:st_kt}
\end{figure}

 All the above results were given in the $s$-domain, i.e. as a function of the variance. In order to go back to the time representation, one needs to solve Eq. (\ref{sdot}) for $t$:
     \begin{equation}\label{ts}
      t(s) = t_\mathrm{i} + \frac{\gamma}{2} \int_{s_\mathrm{i}}^{s} \frac{\mathrm{d}y}{D\gamma - s \bar{\kappa}(y)} ,
    \end{equation}
 which can also be used to obtain the total duration $\Delta t = t_\mathrm{f} - t_\mathrm{i}$.
The time-dependent solutions $\bar{\kappa}(t)={\bar{\kappa}}(s(t))$ and $\kappa(t) = {\kappa}(s(t))$ are depicted in Fig.~\ref{fig:st_kt}, together with the time evolution of the energy of the system, whose integral is the cost functional $F_E$ associated to $\lambda$. The curves corresponds to the same values of $\lambda$ and $\mu$ as in Fig.~\ref{fig:energy_integral}, with total durations $\Delta t = 0.95$, $0.87$ and $0.82$ (left panels, $\mu$ fixed) and $\Delta t = 0.82$, $1.27$ and $1.58$ (right panels,  $\lambda$ fixed). We recall that here the classical relaxation time is equal to $\tau_{\rm rel}= \sqrt{m/\kappa_\mathrm{f}} =\omega_\mathrm{f}^{-1} = 2$, and therefore $\Delta t < \tau_{\rm rel}$ everywhere.

When  the Lagrange multiplier $\lambda$, corresponding to the functional $F_E = \int_{t_\mathrm{i}}^{t_\mathrm{i}} \mathrm{d}t\, E(t)$, is increased, then the energy integral decreases, as seen in the inset of Fig. \ref{fig:st_kt}(e). This is natural, as the role of the $\lambda$-term is precisely to limit the value of the cumulative energy over time. However, we also observe that
the time duration $\Delta t$ decreases for increasing $\lambda$, while the opposite occurred  for classical systems \cite{Rosales-Cabara2020a}. This may seem surprising, as increasing $\lambda$ should reduce the value of the functional $F_E$ while increasing $\Delta t$ (energy-time trade-off).
But, in contrast to the classical overdamped case, here there are two Lagrange multipliers, so the trade-off is actually among \emph{three} functionals, $F_E$, $\Delta t$, and $G$, which makes the whole situation more complex.

Conversely, the time duration  increases as $\mu$ increases, which is more in line with a trade-off between $G$ and  $\Delta t$.
At the same time, the cumulative energy over time increases with increasing $\mu$, as seen in the inset of Fig. \ref{fig:st_kt}(f).

For all cases, the behavior of the variance appears very similar when plotted against time normalized to the duration $\Delta t$.
As expected, $s(t)$ is strictly increasing in time and has vanishing derivatives close to the initial and final times. The energy $E(t)$  follows closely the evolution of $\kappa(t)$, which indicates that the potential energy is the preponderant contribution to the total energy of the system. It is important to note that, when $\mu$ becomes small enough, the energy might become negative, as seen in Fig.~\ref{fig:st_kt}(f), although its time integral remains strictly positive, ensuring that the functional $F_E$ is positive.

It is interesting to compare our results with other protocols, such as the ones that minimizes the work done on the system, proposed by Chen et al. \cite{Chen2010}. This protocol is obtained from the Ermakov equation \eqref{ermakov} rewritten in terms of the scale factor $q(t)=\sqrt{s(t)/s_\mathrm{i}}$, which reads as: $\ddot{q} + \frac{\kappa(t)}{m} q = \kappa_\mathrm{i}/q^3$. This equation can be inverted to obtain the stiffness as a function of the scale factor: $\kappa =\kappa_\mathrm{i}/q^4 -  m\ddot{q}/q$. The protocol is then obtained by imposing a suitable temporal profile for $q(t)$. The authors of \cite{Chen2010} chose a polynomial of fifth degree:  $q(t) = (a-1)(6T^5 -15T^4 +10T^3 +1)$, where $T=t/t_\mathrm{f}$ and $a=(\kappa_\mathrm{i}/\kappa_\mathrm{f})^{1/4}$, so that $q(t_\mathrm{i})=1$, $q(t_\mathrm{f})=a$, and its first and second derivatives are zero at the initial and final times. We recall that imposing such boundary conditions is sufficient to minimize the external work, which becomes equal to the difference between the final and initial energies.

\begin{figure}[htbp]
      \centering
      \includegraphics[width=0.4\paperwidth]{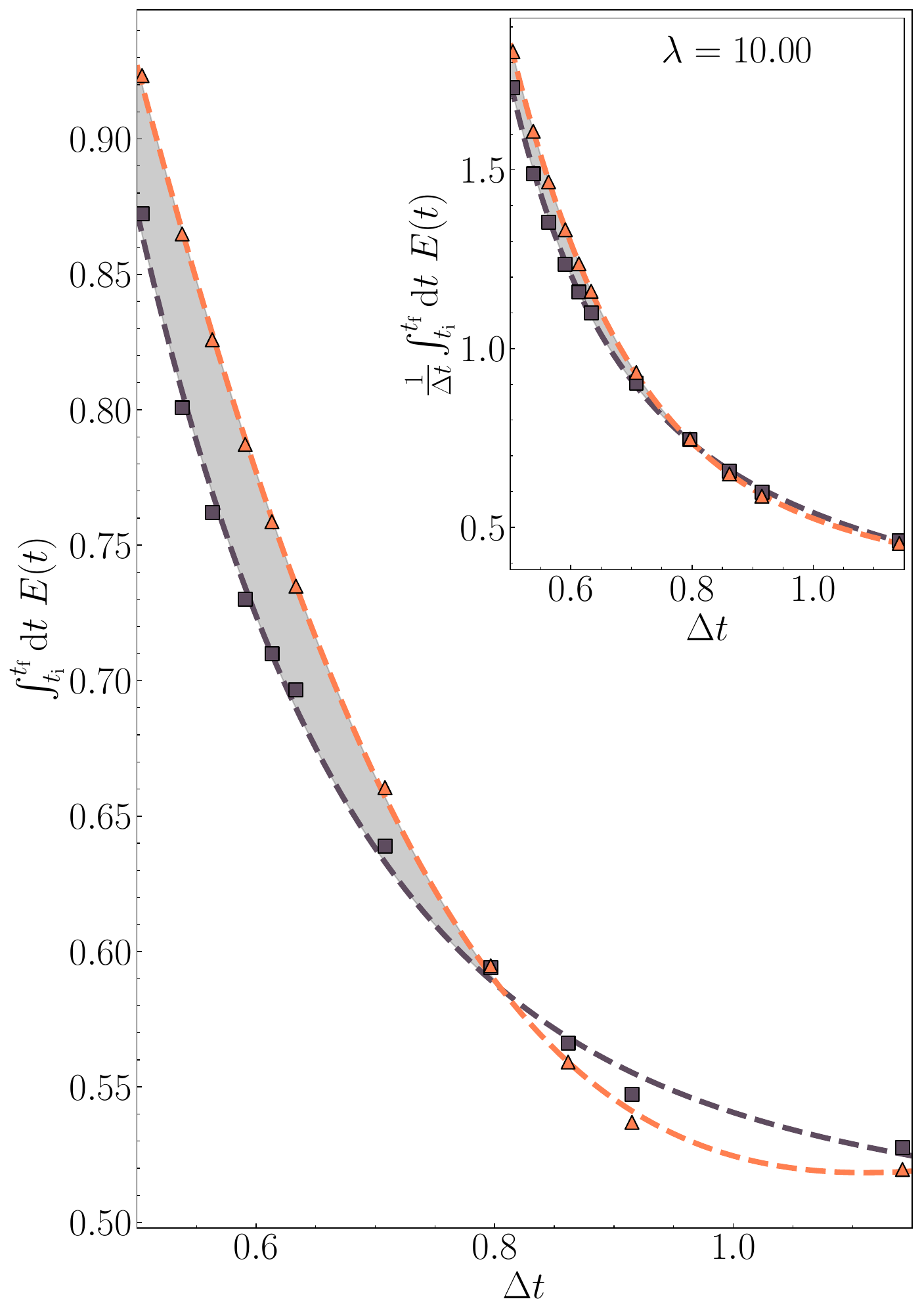}
      \caption{Cumulative energy of the system over time $\int_{t_\mathrm{i}}^{t_\mathrm{f}} \mathrm{d}t\, E(t)$ as a function of the total time duration $\Delta t$ of the protocol. The black squares corresponds to the optimal protocol, solution of the Euler-Lagrange equation \eqref{euler_lagrange_energy}, while the orange triangles correspond to the polynomial protocol of Ref. \cite{Chen2010}. In the gray shaded region, the optimal protocols display lower cumulative energy than the polynomial ones. The different times durations are obtained by varying the Lagrange multiplier $\mu$, while keeping $\lambda=10$ fixed.
      The dashed lines are empirical fits to the numerical data.
      The inset shows the same plots for the time-averaged energy, given by the integral of the energy divided by the time duration.
 }
 \label{fig:optimality_energy}
    \end{figure}

To compare Chen's polynomial protocol to our optimal protocol, we can fix a value of $\lambda$, for instance $\lambda = 10$, and vary the value of $\mu$. For each protocol, we compute the energy integral $F_E = \int_{t_\mathrm{i}}^{t_\mathrm{f}} E(t)  \mathrm{d}t$, as in Eq. \eqref{Fintegral}, and the time duration $\Delta t$, Eq. \eqref{tf}. The results are shown in Fig.~\ref{fig:optimality_energy} for both our protocol (black squares) and Chen's (triangles). It appears that our protocol performs better, in terms of the integral of the energy, than the polynomial protocol for small values of $\Delta t$ (gray area in Fig. \ref{fig:optimality_energy}). For larger $\Delta t$, the opposite is true, which can be explained by noticing that the long-duration regime corresponds to large values of $\mu$. Indeed, the total functional  $J = \Delta t + \lambda F + \mu G$, see Eq. \eqref{J}, can be minimized  by minimizing either $F$ or $G$. If the Lagrange multiplier $\mu$ is large, then $G$ dominates, and the variational procedure will end up minimizing $J$ by essentially minimizing $G$ instead of $F$. However, this is not a significant constraint for our purposes, since our aim is to minimize the energy integral for short durations, which correspond to small values of $\mu$.
In the inset of the same figure \ref{fig:optimality_energy}, we also plot the cumulative energy over time divided by $\Delta t$, which represents the \emph{average} energy of the system. Again, our optimal protocol is the one that minimizes the averaged energy for short durations.

\subsection{Dynamical phase as cost function}\label{sec:dynamical-cost}

As a second example, we consider the following functional to be optimized: $F_{\alpha} =  \int_{t_\mathrm{i}}^{t_\mathrm{f}}\mathrm{d}t~\alpha^2(t)$,
where $\alpha(t)$ is the dynamical phase given in Eq. (\ref{psi_gaussian}). Before carrying out the optimization procedure, we provide some justification about the importance and meaning of such functional.
According to Eq. (\ref{alpha}), $\alpha(t)$  is proportional to the time derivative of the variance. During an adiabatic process, ${\dot s}(t) \approx 0$, because the process is infinitely slow, and therefore $\alpha(t) \approx 0$. Hence, the integral  $F_{\alpha} =  \int_{t_\mathrm{i}}^{t_\mathrm{f}}\mathrm{d}t~\alpha^2(t)$ represents the departure from adiabaticity, and by minimizing such quantity we therefore minimize the ``distance" of the optimal protocol from an adiabatic one.

\begin{figure}[htbp]
      \centering
        \includegraphics[width=0.7\paperwidth]{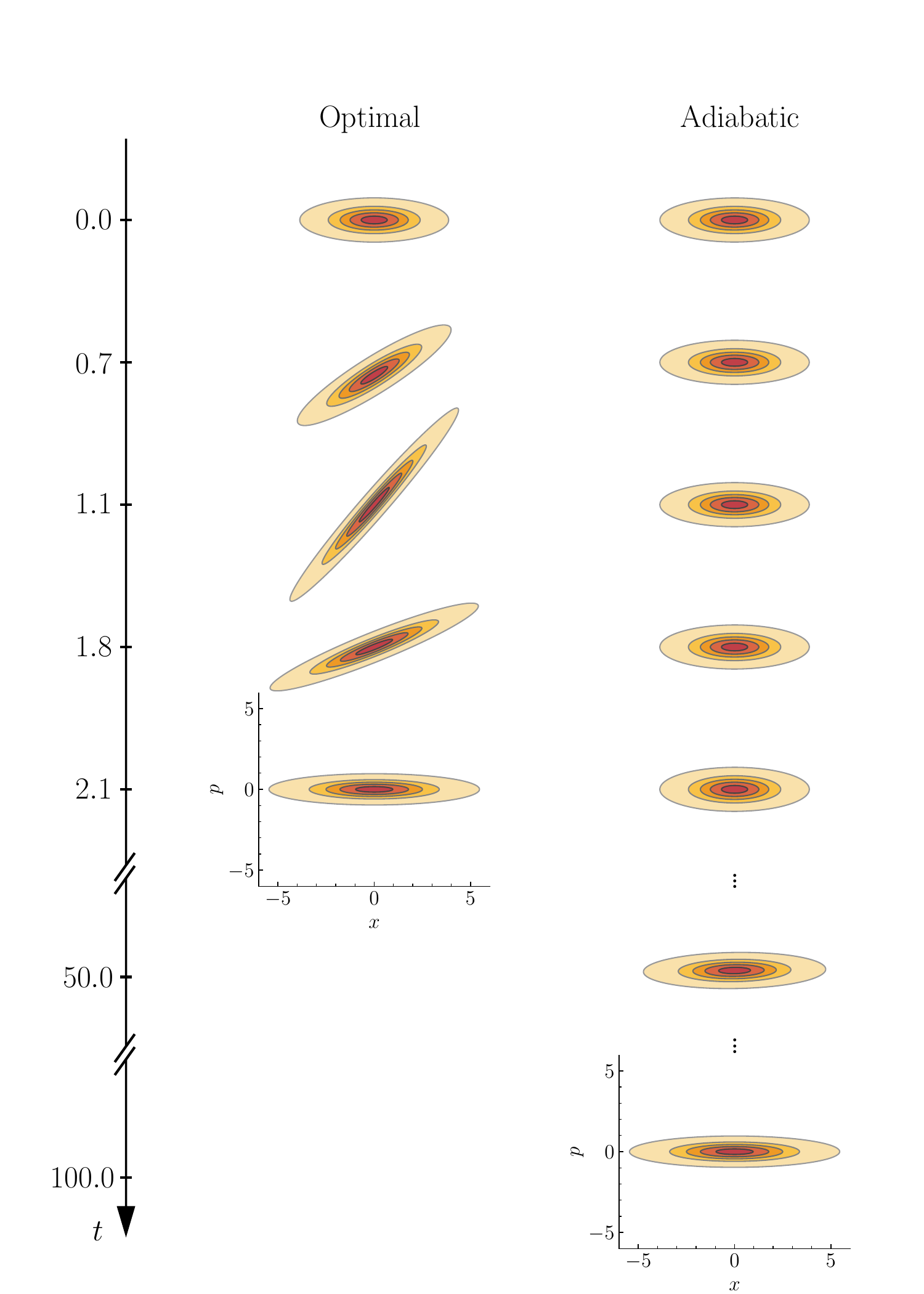}
        \caption{Wigner functions $W(x,p,t)$ at several instants indicated on the vertical time axis. The left panels refer to the phase-optimal protocol (with $\lambda=\mu=1$), while the right panels correspond to an adiabatic protocol. The variance grows from $s_\mathrm{i}=1$ to $s_\mathrm{f}=2$. The Wigner function of the adiabatic protocol remains symmetric with respect to both the $x$ and $p$ axes, and slowly reduces its width in $p$ while increasing its width in $x$.
       In contrast, the phase-optimal Wigner function first becomes elongated along the straight line $p=2\alpha\hbar x$, before  reaching the same final state in a time much shorter than the adiabatic protocol. }
        \label{fig:phasespace}
    \end{figure}

A visual way to represent the dynamical phase is to consider the Wigner function $W(x,p,t)$ \cite{Wigner1932a} corresponding to the Gaussian wave packet of Eq. (\ref{psi_gaussian}).
The Wigner function is a quantum pseudo-probability density in the phase space $(x,p)$.
As the wavefunction is the exponential of a quadratic polynomial, its Wigner function is non-negative, and can be written as \cite{Hudson1974}
\begin{equation}
W(x,p,t) = A \exp\left( -\frac{x^2}{2s(t)} - \frac{2s(t)}{\hbar^2} \left[p-2\alpha(t) \hbar x\right]^2 \right) ,
\label{Wigner}
\end{equation}
where $A$ is a normalization constant  ensuring that $\int_{\mathbb{R}^2}  W \,dx\,dp = 1$. Note that $W(x,p,t)$ peaks around the straight line in phase space $p_0(x,t) = 2\alpha(t) \hbar x = \hbar \partial_x S(x,t)$, where $S(x,t)$ is the total phase of the Gaussian wavefunction (\ref{psi_gaussian}).
When $\alpha=0$, the Wigner function is symmetric with respect to both the position and momentum axes; when $\alpha \neq 0$, it is tilted of an angle $\theta$ such that $\tan \theta = 2\alpha\hbar/(m\omega_\mathrm{i})$ (obtained by expressing $p$ and $x$ in our normalized units).
During an adiabatic protocol, $W(x,p,t)$ remains symmetric and only changes its aspect ratio. For instance, during an expansion ($s_\mathrm{f} > s_\mathrm{i}$), it becomes wider in $x$ and narrower in $p$.
Instead, during a faster-than-adiabatic process,  $W$ first gets  tilted of an angle $\theta$, then expands by increasing its
spatial variance, and finally recovers a symmetric shape with $\alpha=0$.
This is illustrated in Fig. \ref{fig:phasespace}, where we show the phase space portraits for the optimal and adiabatic protocols at several instants in time. It is clear that the acceleration in the optimal protocol is achieved by tilting the Wigner function of a certain angle before coming back to a symmetric configuration at the end of the protocol.

Turning back to the minimization procedure, the $F_{\alpha}$ functional is given by, after changing variable from $t$ to $s$ and expressing the integrand in terms of ${\bar{\kappa}}(s)$,
 \begin{equation}
      F_{\alpha}[s,{\bar{\kappa}}] = \int_{t_\mathrm{i}}^{t_\mathrm{f}}\mathrm{d}t~\alpha^2(t) = \frac{m^2}{8\gamma\hbar^2}\int_{s_\mathrm{i}}^{s_\mathrm{f}}\mathrm{d}s~\frac{D\gamma-s{\bar{\kappa}}(s)}{s^2} ,
\end{equation}
where we have used Eq. \eqref{sdot}.
The total Lagrangian is
 \begin{equation}
      L[s,{\bar{\kappa}},{\bar{\kappa}}'] = \frac{\gamma}{D\gamma-s{\bar{\kappa}}(s)} + \lambda \frac{m^2}{8\gamma\hbar^2}\frac{D\gamma-s{\bar{\kappa}}(s)}{s^2} + \mu {\bar{\kappa}}'^2,
  \end{equation}
 leading to the following Euler-Lagrange equation:
    \begin{equation}\label{euler-lagrange_alpha}
      2\mu\,{\bar{\kappa}}'' = \frac{\gamma s}{(D\gamma - s{\bar{\kappa}})^2} - \frac{m^2\lambda}{8\gamma\hbar^2 s}.
    \end{equation}
Again, this is a  second-order non-linear ordinary differential equation, whose initial and final conditions ${\bar \kappa}_\mathrm{i,f}$ can be imposed to ensure the equilibrium conditions.

\begin{figure}[htbp]
      \centering
        \includegraphics[width=0.4\paperwidth]{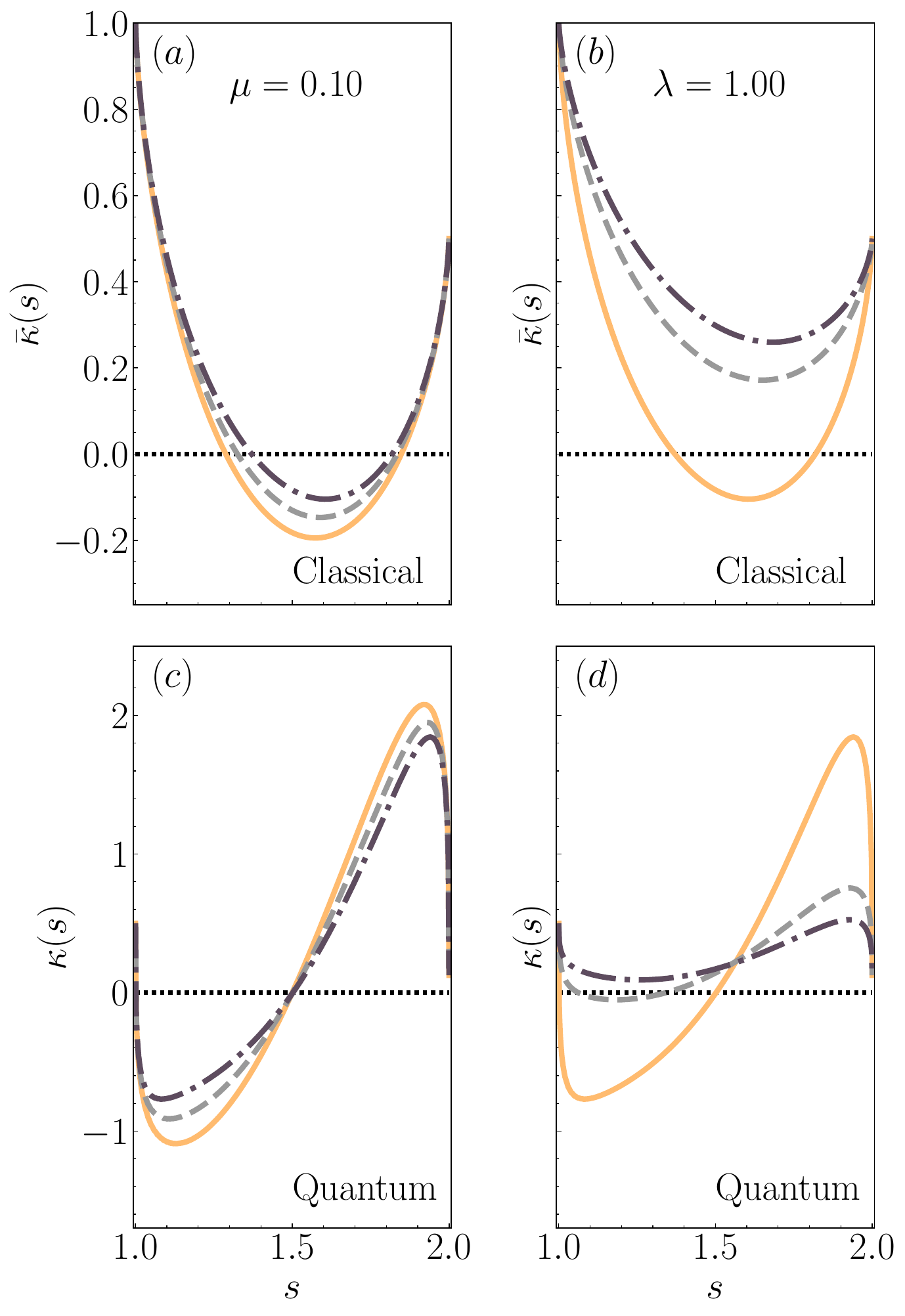}
\caption{Phase-optimal protocols in the $s$-domain, where the variance increases from $s_\mathrm{i} =1$ to $s_\mathrm{f} = 2$.
      Left panels: Classical protocols $\hbk(s)$ (a) and quantum protocols $\kappa(s)$ (c),
      for a fixed value of $\mu = 0.10$ and  $\lambda = 0.01$ (solid orange lines), $0.50$ (dashed grey lines) and $1.00$ (dot-dashed black lines).
      Right panels: Classical protocols $\hbk(s)$ (b) and quantum protocols $\kappa(s)$ (d)
      for a fixed value of $\lambda =1$ and for  $\mu = 0.1$ (solid orange lines), $0.50$ (dashed grey lines) and $1.00$ (dot-dashed black lines).
      Note that the quantum protocol can become negative for sufficiently large values of $\lambda$ or sufficiently small values of $\mu$.}
        \label{fig:dynamical_phase}
    \end{figure}

We use the same parameters as in Sec. \ref{sec:energy-cost}, notably $s_\mathrm{i} = 1$ and $s_\mathrm{f} = 2$. In Fig. \ref{fig:dynamical_phase} we represent the classical and quantum protocols for two cases:
(i) a fixed value of $\mu = 0.10$ and $\lambda$ varying  in the range $\lambda \in [0.01, 0.10, 1.00]$ (left panels), and (ii)
a fixed value of $\lambda = 1.00$ and $\mu$  varying  in the range $\mu \in [0.10, 0.50, 1.00]$ (right panels). These correspond
to the following time durations:  $\Delta t = 0.95$, $0.97$, and $1.00$ (fixed $\mu$, left) and $\Delta t = 1.00$, $1.67$, and $2.10$ (fixed $\lambda$, right).
We recall that the classical relaxation time is equal to $\tau_{\rm rel}= \sqrt{m/\kappa_\mathrm{f}} =\omega_\mathrm{f}^{-1} = 2$.
The general behaviors of the protocols look similar to those obtained in Sec. \ref{sec:energy-cost} for the energy cost function, probably because the solutions are somewhat dominated by the $\mu$ term.
A notable difference is that the time duration is less sensitive to the values of the Lagrange multipliers.
Also, the duration increases as $\lambda$ increases, in contrast to what was observed for the energy-optimal protocols of Sec. \ref{sec:energy-cost}.

\begin{figure}[htbp]
      \centering
        \includegraphics[width=0.4\paperwidth]{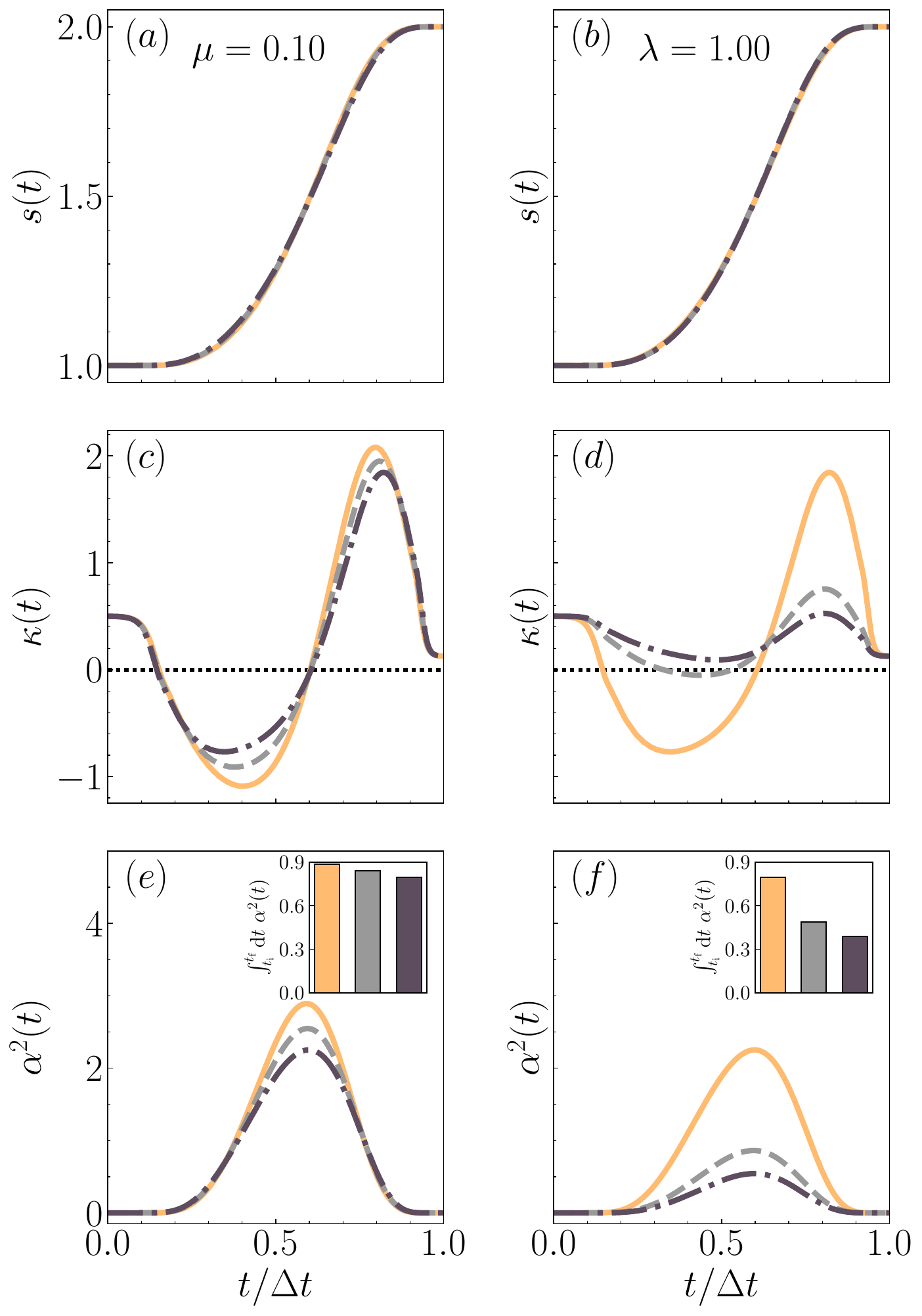}
        \caption{Phase-optimal protocols in the time domain, where the variance increases from $s_\mathrm{i} =1$ to $s_\mathrm{f} = 2$. The time is normalized to the total duration of the protocol $\Delta t$, which is different for each pair of Lagrange multipliers $(\lambda,\mu)$.
      Left panels: variance $s(t)$ (a), quantum protocols $\kappa(t)$ (c), and instantaneous energy $E(t)$ (e), for a fixed value of $\mu = 0.10$ and  $\lambda = 0.01$ (solid orange lines), $0.50$ (dashed grey lines) and $1.00$ (dot-dashed black lines). These values correspond to time durations: $\Delta t = 0.95$, $0.97$ and $1.00$.
      Right panels: variance $s(t)$ (b), quantum protocols $\kappa(t)$ (d), and instantaneous energy $E(t)$ (f), for a fixed value of $\lambda =1$ and  $\mu = 0.1$ (solid orange lines), $0.50$ (dashed grey lines) and $1.00$ (dot-dashed black lines).
      These values correspond to time durations: $\Delta t = 1.00$, $1.67$ and $2.10$.
     The boundary values of the quantum protocols correspond to the conditions: $s_\mathrm{i,f}^2 \kappa_\mathrm{i,f} = mD^2$.
 The insets in (e) and (f) show the integral of $\alpha^2(t)$ over time for the three protocols represented in the same panels.
     Note that the shortest durations correspond to the largest values of the phase $\alpha(t)$.}
      \label{fig:variance_dynamical_phase}
    \end{figure}

The temporal evolution of the variance, the quantum stiffness, and the dynamical phase are depicted in Fig.~\ref{fig:variance_dynamical_phase}, where the time has been normalized to the total duration of each protocol.
In accordance with the above discussion on the Wigner functions, the dynamical phase vanishes at the initial and final times and is maximal around $\Delta t/2$. We also note that the maximum of $\alpha^2(t)$ and its time integral are smaller for the longer time durations of the protocols, i.e., those protocols that are closer to adiabaticity, in agreement with our earlier interpretation of the functional $F_{\alpha}$ as quantifying the departure from adiabaticity.

\begin{figure}[htbp]
      \centering
        \includegraphics[width=0.4\paperwidth]{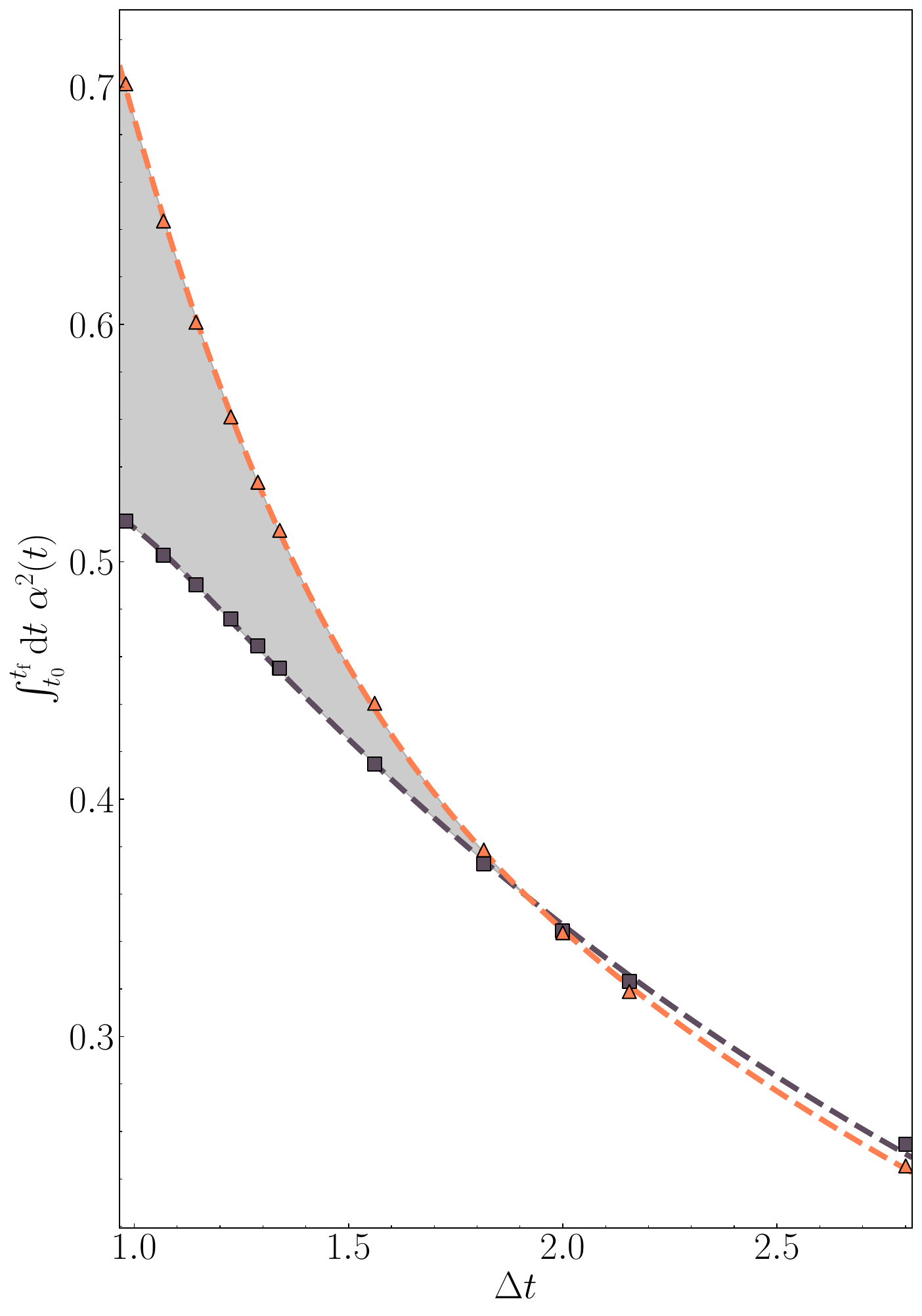}
 \caption{Integral of the square of the dynamical phase $\int_{t_\mathrm{i}}^{t_\mathrm{f}} \mathrm{d}t\, \alpha^2(t)$ as a function of the total time duration $\Delta t$ of the protocol. The black squares corresponds to the optimal protocol, solution of the Euler-Lagrange equation \eqref{euler-lagrange_alpha}, while the orange triangles correspond to the polynomial protocol of Ref. \cite{Chen2010}. In the grey shaded region the optimal protocols perform better than the polynomial ones. The different times durations are obtained by varying the Lagrange multiplier $\mu$ (from $0.001$ to $2.5$), while keeping $\lambda=10$ fixed.
 The dashed lines are empirical fits to the numerical data. }
      \label{fig:optimality_dynamical_phase}
    \end{figure}

In Fig. \ref{fig:optimality_dynamical_phase}, we compare our results to the same polynomial protocol described in Sec. \ref{sec:energy-cost}, by representing the values of the functional $F_{\alpha}$ for different time durations $\Delta t$ for both protocols. Here, we fix $\lambda=10$ and vary $\mu$ from $0.001$ to $2.50$, in order to obtain different time durations for the optimal protocol.
The optimal protocol displays significantly lower values of the cost function $F_{\alpha}$ for short durations, which is the regime of interest. For longer durations, the two protocols behave very similarly in this respect.

The above result is important, inasmuch as it shows that, for the same time duration $\Delta t$, our optimal protocol is closer to adiabaticity than the polynomial protocol of Ref. \cite{Chen2010}.
Both protocols constitute  shortcuts to adiabaticity, but the one we propose here is, in a precise sense, ``adiabatically optimal".

\section{Conclusion} \label{sec:conclusion}

To bring a classical or quantum system from a stationary state to another, the simplest strategy is to vary an external parameter very slowly, i.e. adiabatically. By doing so, the system will be at steady state at each instant of the evolution, but the transition will take an infinite time.
The growing field of research known as ``shortcuts to adiabaticity" tries to accomplish the same transition within a finite duration.

In the present work, we proposed a new strategy to achieve faster-than-adiabatic transitions. The main idea is based on Nelson's representation of quantum mechanics as a classical stochastic process. In the case of a time-dependent harmonic oscillator this quantum-classical analogy is particularly simple and fruitful.  Using Nelson's procedure, the Schr\"odinger equation is rewritten as an overdamped Langevin equation with a linear harmonic force of stiffness $\bar \kappa(t)$, which is related to the stiffness of the quantum oscillator $\kappa(t)$ through Eq. \eqref{quantum_protocol}. Thanks to this mathematical analogy, it is possible to translate the classical protocols developed for an overdamped oscillator into quantum protocols for a system with finite inertia.

In particular, we utilized our experience on optimal classical protocols to devise quantum protocols that minimize both the time duration and some other arbitrary cost function. For instance, the cost function can be the cumulative energy over time.
Even more interestingly, we showed that minimizing the dynamical phase of the wavefunction (again, together with the time duration) amounts to minimizing the distance of the protocol from an adiabatic one.
Hence, we could devise a family of protocol that are ``adiabatically optimal": for a given finite duration $\Delta t$, they are as close as possible to an adiabatic (i.e., infinitely slow) process.

The proposed method is rather versatile, inasmuch as the cost functional to be minimized can be chosen at will. Nevertheless, some functionals may lead to complicated Euler-Lagrange equations, which are difficult to solve numerically. This occurs because the cost functional $F$ that has a physical relevance is expressed in terms of the quantum stiffness $\kappa$ and its time derivative, but must be rewritten in terms of the classical stiffness $\bar \kappa$ before performing the minimization procedure. This can transform relatively a simple functional $F[\kappa, \dot \kappa]$ into a rather complicated functional of
 $\bar \kappa$ and $\dot{\bar \kappa}$.

The original method proposed in this work, although limited here to the ground state of the harmonic potential,  opens up many possible avenues for future research.
For instance, as shown in \cite{Chen2010}, the same procedure also works for transitioning an excited state of the harmonic oscillator $\psi_n(x)$, which transforms according to the scale factor defined earlier as: $q(t)=\sqrt{s(t)/s_\mathrm{i}}$.
In contrast, for anharmonic systems, the Gaussian wavefunction is no longer an exact solution. However, a Gaussian ansatz can be used as an approximate solution, leading to a modified Ermakov equation, similar to Eq. \eqref{ermakov}, which can be used as the basis for a generalization of the present theory. Similarly, one may consider many-body problems in the mean field approximation, either with contact interactions (Gross-Pitaevskii equation for Bose-Einstein condensates) \cite{Manfredi2008BEC,delCampo2011,delCampo2013} or Coulomb interactions (Schr\"odinger-Poisson equations for a quantum electron gas) \cite{Manfredi2009Breather}. These are nonlinear Schr\"odinger equations that are amenable to the Nelson representation utilized in the present work.  For weak coupling, the exact solution can be approximated by a Gaussian wavefunction, leading again to a modified Ermakov equation.
For the Gross-Pitaevskii case, an  interesting goal would be to control the system by modulating the scattering length, which can be done experimentally by varying an external magnetic field.
A further avenue for future research is to extend the present method to the case of an open quantum system in contact with a bath at finite temperature. To do this, it would be necessary to extend Nelson's formalism, for instance by adding a thermal noise to Eq. \eqref{nelson_equation}, for which  several attempts have already been proposed \cite{Ruggiero1985,Kobayashi2011}.

\begin{acknowledgments}
We thank Dr. Cyriaque Genet for many useful discussions.
R. G. acknowledges support from the Mark Ratner Institute for Single Molecule Chemistry at Tel Aviv University.
This work was funded by the French National Research Agency (ANR) through the Programme d'Investissement d'Avenir under contract ANR-11-LABX-0058-NIE and ANR-17-EURE-0024 within the Investissement d’Avenir program ANR-10-IDEX-0002-02.
\end{acknowledgments}
\newpage
\clearpage

\appendix

\section{Classical work-optimal protocol} \label{annexe}

In order to illustrate the quantum-classical analogy, and to show the importance of adding the functional $G[\hbk']$ to the total functional to be minimized, we propose to study a simple and well-documented case: the classical work-optimal protocol developed in~\cite{Rosales-Cabara2020a}. A Brownian particle is trapped in a harmonic potential whose stiffness $\bk(t)$ can vary in time. The particle is immersed in a fluid of damping coefficient $\gamma$ and thermal diffusion coefficient $D=k_B T/\gamma$, where $T$ is the temperature of the fluid. In Ref. \cite{Rosales-Cabara2020a}, the objective was to find the optimal manner to vary $\bk(t)$ so that both the duration of the transition and the work done on the system are minimal. The position of the Brownian particle follows a Gaussian probability distribution of variance $s(t)$, which obeys Eq.~(\ref{sdot}).
Changing the independent variable from the time $t$ to the variance $s$, we can write the time duration $\Delta t$ as in Eq. (\ref{tf}) and the work done on the system as \cite{Ciliberto2017,Sekimoto1998,Rosales-Cabara2020a}: $W = {1\over 2}\int_{t_\mathrm{i}}^{t_\mathrm{f}} \mathrm{d}t\, \dot {\bar\kappa}(t) \langle x^2 \rangle = -{1\over 2}\int_{s_\mathrm{i}}^{s_\mathrm{f}} \mathrm{d}s~\hbk(s) + {1\over 2}(s_\mathrm{f}\bk_\mathrm{f} - s_\mathrm{i}\bk_\mathrm{i})$.
Hence, the functional to be minimized is
    \begin{equation}
        J[\hbk] = \int_{s_\mathrm{i}}^{s_\mathrm{f}} \mathrm{d}s~\frac{\gamma}{D\gamma - s\hbk(s)} - \lambda \int_{s_\mathrm{i}}^{s_\mathrm{f}} \mathrm{d}s~\hbk(s) ,
        \label{Jclass}
    \end{equation}
with $\lambda$ a Lagrange multiplier. It is straightforward to find the solution of the associated Euler-Lagrange equation \cite{Rosales-Cabara2020a}:
    \begin{equation}
        s\hbk(s) = D\gamma \mp \sqrt{\frac{\gamma s}{\lambda}},
        \label{class-overdamped}
    \end{equation}
where the upper and lower signs correspond to the cases of the compression or expansion, respectively.
Note that, as the Euler-Lagrange equation is purely algebraic, the boundary conditions cannot be fixed at will. Hence, in the classical case, the solution \eqref{class-overdamped} must be supplemented by ``jumps" at the initial and final times  \cite{Rosales-Cabara2020a}.

The associated quantum protocol is obtained  using Eq. (\ref{quantum_protocol_s}), yielding:
    \begin{equation}
        \hk(s) = mD^2 /s^2 ,
        \label{ks-classical}
    \end{equation}
which  is independent of $\lambda$. Surprisingly, this solution coincides with the equilibrium solution \eqref{equilibrium}, which means that it represents an adiabatic process for the quantum oscillator, albeit with a finite duration that can be obtained from Eq. (\ref{tf}): $\Delta t = \sqrt{\gamma \lambda}(\sqrt{s_\mathrm{i}} \pm \sqrt{s_\mathrm{f}} )$.
The variance can be computed solving Eq. (\ref{sdot}), yielding
    \begin{equation}
        s(t) = (\sqrt{s_\mathrm{i}} \pm t/\sqrt{\gamma \lambda})^2,
         \label{s-classical}
    \end{equation}
and the classical and quantum protocols are, respectively,
    \begin{equation}
        \bk(t) = \frac{D\gamma + \sqrt{\gamma s_\mathrm{i}/\lambda} \pm t/\lambda}{( \sqrt{s_\mathrm{i} } \pm t/\sqrt{\gamma \lambda} )^2}\, ; \quad
        \kappa(t) = \frac{mD^2}{(\sqrt{s_\mathrm{i}} \pm t/\sqrt{\gamma\lambda})^4} .
        \label{kappa-classical}
    \end{equation}
Note that, if the time is expressed in units of total time duration $\Delta t$, then $\kappa(t/\Delta t)$ is indeed independent of the Lagrange multiplier $\lambda$.

As mentioned above, the quantum solution is at equilibrium at each instant. However, since the classical equilibrium conditions do not hold, $s_\mathrm{i,f}\bk_\mathrm{i,f}\neq D\gamma$, the time derivative of the variance at the initial and final times is not zero. From the point of view of the classical system this is not a problem, because the overdamped dynamics displays no inertia, so that one can change the stiffness abruptly to bring it to the equilibrium value compatible with $s_\mathrm{f}$ \cite{Rosales-Cabara2020a}.
But for the (inertial) Schr\"odinger equation, if $\dot{s}\neq 0$ at $t=t_\mathrm{f}$, then the system will continue to evolve in time after $t_\mathrm{f}$.

It is therefore necessary to ensure that $D\gamma - s\bk = 0$, both at $t=t_\mathrm{i}$ and $t=t_\mathrm{f}$. In order to do that, the Euler-Lagrangian equation should be a second-order differential equation, instead of an algebraic one as was the case for the functional of Eq. \eqref{Jclass}. This is the reason why one needs to add a second functional of the form $G[\hbk']=\int_{s_\mathrm{i}}^{s_\mathrm{f}}\mathrm{d}s~ \left|\hbk'(s)\right|^2$, associated with the Lagrange multiplier $\mu$, which leads to the following Euler-Lagrange equations
    \begin{equation}
        2\mu\, \hbk''(s) = \frac{\gamma s}{[D\gamma - s\hbk(s)]^2} - \lambda.
        \label{EL-classical}
    \end{equation}
This being a second-order differential equation, the boundary conditions at $t_\mathrm{i}$ and $t_\mathrm{f}$ can be imposed consistently with the requirement that: $s_\mathrm{i,f} \bk_\mathrm{i,f} = D\gamma$.

\begin{figure}
\centering
\includegraphics[width=0.4\paperwidth]{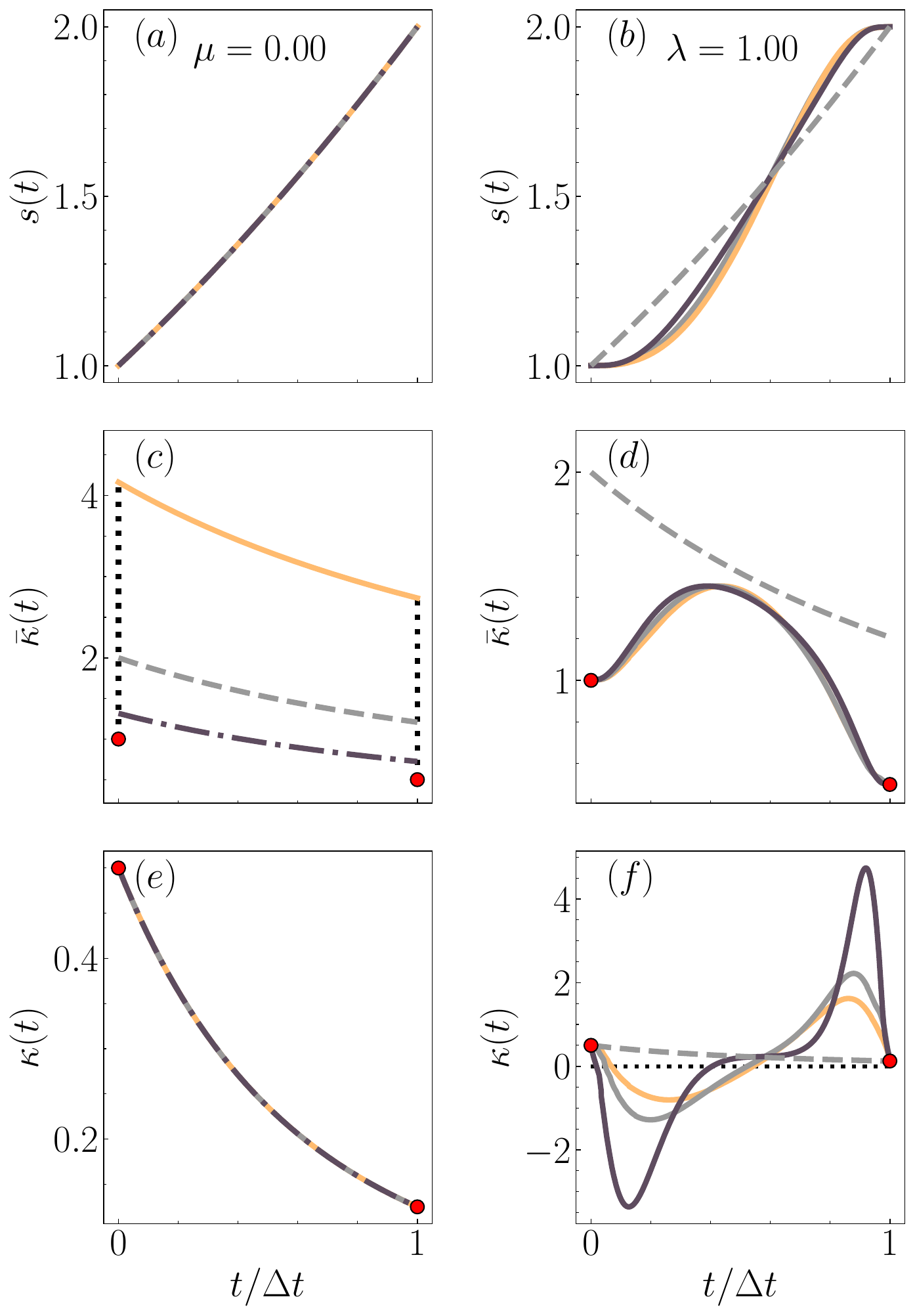}
\caption{Classical work-optimal protocols.
The left panels represent, from top to bottom, the time evolution of the variance (a), the classical protocol (c), and the quantum protocol (e), for the case $\mu=0$, which displays jumps at the initial and final times (the initial and final conditions are represented by red dots).
The different curves are obtained with $\lambda = 0.1$ (orange solid lines), $1.00$ (grey dashed lines) and $10.00$ (black dash-dotted lines). Note that the variance (a) and the quantum protocol (e) do not depend on $\lambda$.
The right panels represent, from top to bottom, the time evolution of the variance (b), the classical protocol (d), and the quantum protocol (f), for fixed $\lambda=1$, and finite values of $\mu$: $\mu = 0.10$ (orange solid lines), $0.05$ (grey solid lines) and $0.01$ (black solid lines). Note that these finite-$\mu$ protocols are continuous and smooth at the initial and final times. For comparison, the dashed grey line represents the classical (discontinuous) protocol with $\mu =0$.}
 \label{fig:apdxA_lam_mu}
\end{figure}

The various results, both for the analytical solution \eqref{class-overdamped}-\eqref{kappa-classical} (with jumps) and the smooth numerical solution of Eq. \eqref{EL-classical}, are presented in Fig. \ref{fig:apdxA_lam_mu} for the variance $s(t)$ (top panels), the classical protocols $\bk(t)$ (middle panels), and the quantum protocols $\kappa(t)$ (bottom panels). In the left panels, we take  $\mu=0$ (no smoothing) and vary $\lambda$ from 0.1 to 10, while the right panels keep  $\lambda=1$ fixed, while $\mu$ varies from 0.01 to 0.1.
It is clear (top left and bottom left panels) that the variance $s(t)$ and the quantum protocol $\kappa(t)$ do not depend on $\lambda$, as suggested by Eqs. \eqref{ks-classical} and \eqref{kappa-classical}.
Instead, the classical protocol depends on $\lambda$, in accordance with Eq. \eqref{class-overdamped}.
It is also evident that the classical protocol displays discontinuities at the initial and final times (which disappear in the adiabatic limit $\lambda\to\infty$), while the quantum protocol does not.

In the protocols with $\mu >0$ (right panels of Fig. \ref{fig:apdxA_lam_mu}), the variance varies smoothly at the initial and final times, as expected, and the classical protocols display no discontinuities at the boundaries. Hence, the equilibrium conditions are fulfilled and the system's variance will remain at its final value at the end of the transition.
Finally, we note that the quantum protocol develops large spikes near $t_\mathrm{i}$ and $t_\mathrm{f}$ for small values of $\mu$. Hence, although it must converge to the non-smoothed one for $\mu \to 0$, it appears to do so in a singular way, displaying large positive and negative spikes at the boundaries.

\bibliography{article_protocol}
\end{document}